\documentclass[12pt]{article}

\usepackage{enumitem,amssymb}
\usepackage{ragged2e}
\usepackage{graphicx}
\usepackage{comment}
\usepackage[utf8]{inputenc}
\usepackage{charter}
\usepackage{fullpage}
\usepackage{hyperref}
\usepackage{graphicx}
\usepackage{lipsum}
\usepackage[sort&compress,numbers]{natbib}
\usepackage{hyperref}
\hypersetup{hidelinks}
\usepackage[total={6.5in,9in},top=1in,headsep=0.1in,headheight=1in]{geometry}
\usepackage[utf8]{inputenc}

\usepackage{aas_macros}

\newcommand\snowmass{
\begin{center}
  \rule[-0.2in]{\hsize}{0.01in}\\
  \rule{\hsize}{0.01in}\\
  \vskip 0.1in
  Submitted to the Proceedings of the US Community Study\\ 
  on the Future of Particle Physics (Snowmass 2021)\\
  \rule{\hsize}{0.01in}\\
  \rule[+0.2in]{\hsize}{0.01in}\\[-2em]
\end{center}
}

\usepackage[firstpage=true]{background}
\backgroundsetup{contents={\parbox{6.5in}{\snowmass}}, scale=1,placement=top,opacity=1,color=black,position={3.25in,1.2in}}

\usepackage{fancyhdr}
\fancypagestyle{plain}{
  
  \fancyhf{}
  \fancyhead[C]{}
  \fancyfoot[C]{\thepage}
}

\fancypagestyle{empty}{
  \fancyhf{}
  \fancyhead[C]{{\it Snowmass2021 Cosmic Frontier White Paper}}
  \fancyfoot[C]{\thepage}
}
\title{Snowmass2021 Cosmic Frontier White Paper:
\\Fundamental Physics and Beyond the Standard Model}
\date{}

\usepackage{authblk}

\author[1]{Emanuele Berti}
\author[2,3]{Vitor Cardoso}
\author[4]{Zolt\'an Haiman}
\author[5]{Daniel E. Holz}
\author[6]{Emil Mottola}
\author[7]{Suvodip Mukherjee}
\author[8,9,10]{Bangalore Sathyaprakash}
\author[11,12]{Xavier Siemens}
\author[13]{Nicol\'as Yunes}

\affil[1]{Department of Physics and Astronomy, Johns Hopkins University, 3400 N. Charles Street, Baltimore, Maryland, 21218, USA}
\affil[2]{Niels Bohr International Academy, Niels Bohr Institute, Blegdamsvej 17, 2100 Copenhagen, Denmark}
\affil[3]{CENTRA, Instituto Superior T\'ecnico, Universidade de Lisboa,
Avenida Rovisco Pais 1, 1049-001 Lisboa, Portugal}
\affil[4]{Department of Astronomy, Columbia University, New York, NY 10027, USA}
\affil[5]{Department of Astronomy and Astrophysics, Department of Physics, Enrico Fermi Institute, and Kavli Institute for Cosmological Physics, University of Chicago, 5640 South Ellis Ave., Chicago, IL 60637}
\affil[6]{Department of Physics and Astronomy, University of New Mexico
Albuquerque, New Mexico 87131, USA}
\affil[7]{Perimeter Institute for Theoretical Physics, 31 Caroline Street N., Waterloo, Ontario, N2L 2Y5, Canada}
\affil[8]{Institute for Gravitation and the Cosmos, Department of Physics,
Pennsylvania State University, University Park, PA, 16802, USA}
\affil[9]{Department of Astronomy \& Astrophysics, Pennsylvania State University, University Park, PA, 16802, USA}
\affil[10]{School of Physics and Astronomy, Cardiff University, Cardiff, UK, CF24 3AA}
\affil[11]{Department of Physics, Oregon State University, Corvallis, OR 97331, USA}
\affil[12]{Center for Gravitation, Cosmology and Astrophysics, University of Wisconsin–Milwaukee, P.O. Box 413, Milwaukee WI, 53201, USA}
\affil[13]{Illinois  Center  for  Advanced  Studies  of  the  Universe and Department of Physics, University of Illinois at Urbana-Champaign, Urbana, Illinois 61801, United States}

\begin{document}

\maketitle

\begin{abstract}
Gravitational wave detectors are formidable tools to explore strong-field gravity, especially black holes and neutron stars. These compact objects are extraordinarily efficient at producing electromagnetic and gravitational radiation. As such, they are ideal laboratories for fundamental physics and have an immense discovery potential. The detection of black hole binaries by third-generation Earth-based detectors, space-based detectors and pulsar timing arrays will provide exquisite tests of general relativity. Loud ``golden'' events and extreme mass-ratio inspirals can strengthen the observational evidence for horizons by mapping the exterior spacetime geometry, inform us on possible near-horizon modifications, and perhaps reveal a breakdown of Einstein’s gravity. Measurements of the black-hole spin distribution and continuous gravitational-wave searches can turn black holes into efficient detectors of ultralight bosons across ten or more orders of magnitude in mass. A precise monitoring of the phase of inspiralling binaries can constrain the existence of additional propagating fields and characterize the environment in which the binaries live, bounding the local dark matter density and properties. Gravitational waves from compact binaries will probe general relativity and fundamental physics in previously inaccessible regimes, and allow us to address fundamental issues in our current understanding of the cosmos.
\end{abstract}

\clearpage

\tableofcontents

\clearpage
\section{Introduction}

Over the past century, advances in our understanding of the Universe have also uncovered new puzzles yet to be solved. What is the nature of BH horizons and their role in the information loss problem, or in censoring singularities? What exactly is dark matter (DM)? Is dark energy simply a small cosmological constant, a ``strange'' fluid or something else, possibly explained at a fundamental level via zero-point quantum fluctuations? Is general relativity (GR) the most accurate description of spacetime, even for the most energetic and gravitationally indomitable events in Nature? Is there a unified theory of fundamental interactions, and can we look for its footprints in the early stages of the cosmos?

Gravitational-wave (GW) and electromagnetic (EM) observations of extremely energetic phenomena in strong gravitational fields have the potential to answer these fundamental questions. The first direct detection of GWs in 2015~\cite{TheLIGOScientific:2016qqj,TheLIGOScientific:2016wfe,TheLIGOScientific:2016pea,Abbott:2016nmj} by the advanced LIGO (aLIGO) detectors~\cite{Abramovici:1992ah,Smith:2009bx} inaugurated the era of GW astrophysics. This observation revealed that two black holes (BHs), each with mass $\gtrsim 30~M_\odot$, collided at about half the speed of light to form another, more massive BH~\cite{TheLIGOScientific:2016qqj}. The GWs detected were consistent with GR, carrying three solar masses in energy and for an instant ``outshining'' all stars put together. Since then we have observed $\sim 90$ mergers, turning GW astronomy into a new tool to study astrophysical populations and test GR in ways that were previously impossible~\cite{Abbott:2016nmj,TheLIGOScientific:2016pea,Abbott:2017vtc,LIGOScientific:2018mvr,Abbott:2020niy}. 

Simultaneously, pulsar timing arrays (PTAs) are at the very cusp of making the first detection of the GW background generated by supermassive BH binaries~\cite{Antoniadis:2022pcn}.  They operate on astrophysical scales, and each ultra-stable pulsar in the array works as a clock that can be timed to nano-second precision with radio telescopes. By combining the times of arrival from multiple pulsars, one can search for correlations that signal the presence of an astrophysical background of GWs.

This is just the beginning. In the mid 2020s, hundreds of GW observations are expected as aLIGO~\cite{LIGOScientific:2014pky} and Virgo~\cite{VIRGO:2014yos} are upgraded in sensitivity, and eventually joined by the Japanese KAGRA detector~\cite{KAGRA:2020tym} and the Indian LIGO-India interferometer~\cite{Saleem:2021iwi}.  In the mid 2030s, these ground-based detectors will be joined by the first space-based GW mission, L
ISA~\cite{LISA:2017pwj}, which for the first time will detect GWs in the mHz band. The next generation of ground-based detectors (Cosmic Explorer~\cite{Evans:2021gyd, Hall:2019xmm, Reitze:2019iox, Borhanian:2022czq} and the Einstein Telescope~\cite{Punturo:2010zz, Sathyaprakash:2012jk, Maggiore:2019uih}) planned for the 2030s), will observe thousands of sources with larger signal-to-noise ratios (SNRs): in fact, they would enable the observation of pretty much \textit{all} stellar-mass binary BH mergers in the Universe.

Ground- and space-based detectors, along with PTAs, produce a complementary scan of the sky, as they operate at very different frequencies. Ground-based detectors operate at relatively high frequencies~\cite{Abramovici:1992ah}; second-generation detectors are sensitive above $\sim 10$ Hz, and only third-generation (3G) detectors could reach frequencies $\gtrsim 1$ Hz.  Consequently, ground based instruments can detect GWs emitted by binaries with masses $\lesssim 10^{2} M_{\odot}$, the events lasting less than a second in band and with SNRs of $\sim 10$--$100$, thus probing the ``local'' Universe. LISA, on the other hand, will operate at lower frequencies (between $\sim 10^{-5}$ and $\sim 10^{-2}$ Hz)~\cite{AmaroSeoane:2012je}, where the source populations are much richer (including the merger of supermassive BHs in major galaxy mergers), and events can last months to years with SNRs in the hundreds to thousands, probing a much larger volume of the Universe. Therefore, LISA will measure tens to hundreds of thousands of GW cycles from massive BH inspirals, encoding rich information from which to draw exquisitely precise astrophysical conclusions and perform stringent tests of GR in extreme gravity. PTAs operate at even lower frequencies, with sensitivity between $10^{-8}$ Hz and $10^{-6}$ Hz (see e.g.~\cite{Antoniadis:2022pcn,NANOGrav:2021ini,NANOGrav:2021flc,NANOGrav:2020spf,NANOGrav:2020bcs}). Therefore, PTAs can detect GWs emitted by the very heaviest BH binaries with masses $\gtrsim 10^{7} M_{\odot}$, thus probing the Universe on large scales.

The combination of GW observatories has an unprecedented potential to answer fundamental questions about the Universe in the extreme gravity regime~\cite{Sathyaprakash:2009xs, Gair:2012nm,Yagi:2016jml,Berti:2018cxi,Barack:2018yly,Berti:2018vdi,Berti:2019xgr,Cardoso:2019rvt,Barausse:2020rsu}, but this potential is greatly enhanced through coincident EM observations. The merger of neutron stars and the coalescence of a neutron star with a BH are believed to produce also a short burst of gamma rays. Indeed, GWs from the first binary neutron star inspiral detected by aLIGO and Virgo in 2017 were accompanied by a short gamma ray burst~\cite{LIGOScientific:2017zic} and by EM radiation in other frequency bands~\cite{LIGOScientific:2017ync}. Together, this multi-messenger observation allowed for the precise localization of the source and the most accurate ever measurement of the propagation speed of GWs~\cite{Creminelli:2017sry,Ezquiaga:2017ekz,Baker:2017hug}.  Binary neutron star events are not the only ones expected to be accompanied by an EM counterpart. The coalescence of supermassive BHs is also believed to emit electromagnetically, if the BHs possess accretion disks that change dynamically due to the merger processes~\cite{Haiman:2019xme,Bogdanovic:2021aav}.

In this White Paper, we will explain what these GW and multi-messenger observations can do for the field of particle physics in the next decade. This White Paper is an extension of five letters of intent (``Fundamental Physics with Gravitational Wave Detectors''~\cite{FPLOI}, ``Multi-messenger Probes of Cosmology and Fundamental Physics using Gravitational Waves''~\cite{MMALOI}, ``Fundamental Physics with Pulsar Timing Arrays''~\cite{PTALOI}, ``Physical Effects of Nonlocally Coherent Quantum Gravity''~\cite{QGLOI}, and ``Searching for Scalar Gravitational Waves in Neutron Star Binary Mergers''~\cite{SGWLOI}) submitted to the Snowmass 2021 Cosmic Frontier process.

Many other White Papers submitted to Snowmass 2021 highlight the central role of GW astronomy for fundamental physics, gravitational physics and cosmology. A partial list includes the White Papers
``Detection of Early-Universe Gravitational Wave Signatures and Fundamental Physics''~\cite{Caldwell:2022qsj},
``Cosmology Intertwined: A Review of the Particle Physics, Astrophysics, and Cosmology Associated with the Cosmological Tensions and Anomalies''~\cite{Abdalla:2022yfr},
``Dark Matter In Extreme Astrophysical Environments''~\cite{Baryakhtar:2022obg},
and
``Probing Dark Matter with Small-scale Astrophysical Observations''~\cite{Brito:2022lmd}.
Other White Papers focus on the numerical relativity and experimental developments necessary to realize this vision
(``Numerical Relativity for Next-generation Gravitational-wave Probes of Fundamental Physics''~\cite{Foucart:2022iwu},
``Future Gravitational-Wave Detector Facilities''~\cite{Ballmer:2022uxx})
and on the important role of multimessenger observations: see e.g.
``Synergies between Dark Matter Searches and Multiwavelength/Multimessenger Astrophysics''~\cite{Ando:2022kzd},
``Multi-Experiment Probes for Dark Energy -- Transients''~\cite{Kim:2022iud},
``Observational Facilities to Study Dark Matter''~\cite{Chakrabarti:2022cbu},
``Astrophysical and Cosmological Probes of Dark Matter''~\cite{Boddy:2022knd}.

\begin{figure}
\centering
\includegraphics[width=\textwidth]{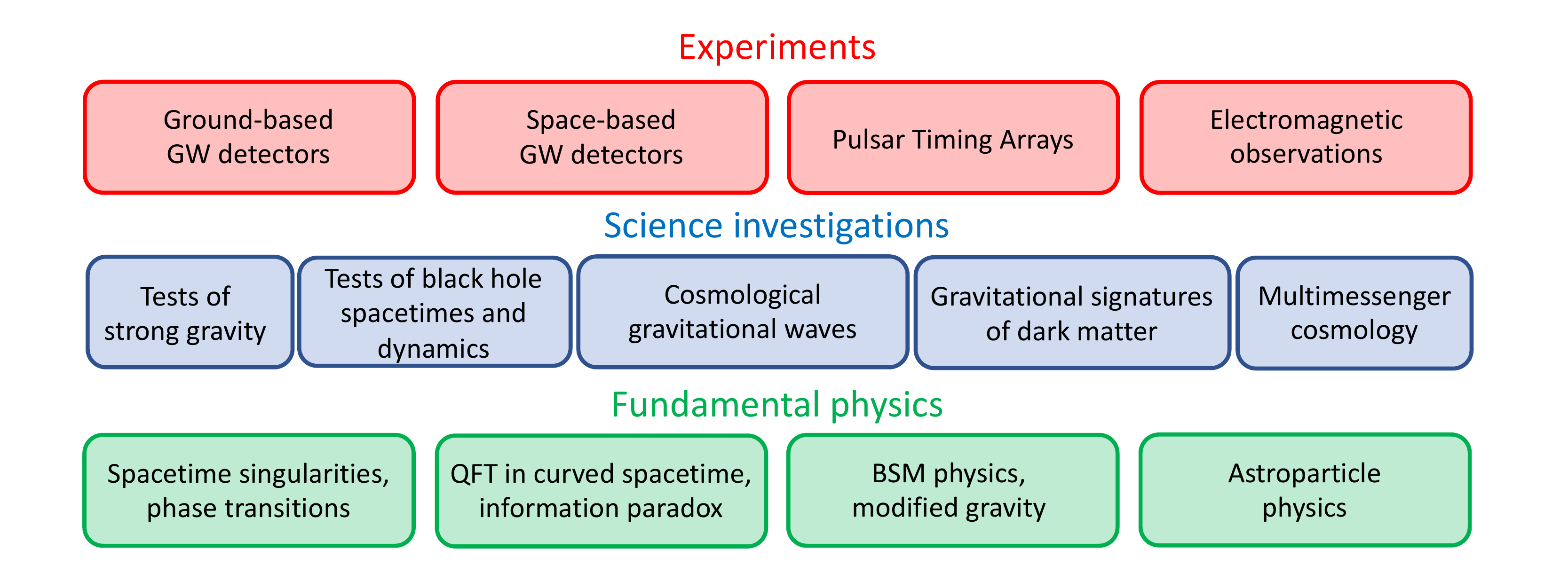}
\caption{Schematic illustration showing how the science investigations addressed in this White Paper (in blue) address fundamental open issues in theoretical physics (in green), and their connection with GW and EM observations (in red).
  \label{fig:flowchart}
}
\end{figure}

The present White Paper is organized through a ``science-first'' approach (see Fig.~\ref{fig:flowchart}), dividing its content into: tests of strong gravity (Sec.~\ref{sec:MG}); BH horizons, quantum gravity, and the information paradox (Sec.~\ref{sec:BHs}); gravitational signatures of DM (Sec.~\ref{sec:DM}), GW and multimessenger cosmology (Sec.~\ref{sec:MMA}); and cosmological GWs (Sec.~\ref{sec:cosmo}).

\section{Tests of strong-field gravity}
\label{sec:MG}

Although Einstein's theory has passed a plethora of tests~\cite{Will:2005va,Yunes:2013dva,Berti:2015itd,Cardoso:2016ryw}, GWs have an unprecedented potential to carry out \emph{precision tests} in the extreme gravity regime~\cite{Gair:2012nm,Yagi:2016jml,Berti:2018cxi,Barack:2018yly,Berti:2018vdi,Berti:2019xgr,Barausse:2020rsu}. Why are such tests valuable? Several fields in theoretical physics, including particle physics, have explored the possibility of modifying GR~\cite{Joyce:2014kja} both for observational reasons (e.g.~to explain the late-time acceleration of the Universe~\cite{SupernovaSearchTeam:1998fmf,SupernovaCosmologyProject:1998vns}, or galaxy rotation curves~\cite{Sofue:2000jx,Bertone:2016nfn}) and for theoretical reasons (e.g.~the incompatibility of quantum mechanics and GR, or the conceptual issues related to BHs discussed in the next section). Some of these modified theories pass all binary pulsar, cosmological and Solar System tests~\cite{Yunes:2013dva,Berti:2015itd,Yagi:2016jml,Barack:2018yly,Berti:2018cxi,Berti:2018vdi,Berti:2019xgr,Barausse:2020rsu,Giannotti:2008cv,Mottola:2010gp,Mottola:2016mpl}, and yet they introduce modifications to gravity in the extreme regime, where GWs can constrain them. This extreme gravity regime, where gravitational fields are very large and change dynamically during the observation time, was not accessible before the dawn of GW astrophysics.

Modified theories of gravity can, in general terms, be grouped into two large classes: those that induce infrared (large scale) modifications, and those that induce ultraviolet (small scale) modifications. Members of the first class typically possess some type of screening mechanism, like the Vainshtein mechanism~\cite{Vainshtein:1972sx,Babichev:2013usa,Babichev:2013pfa,Chkareuli:2011te}. They include theories like massive gravity~\cite{deRham:2014zqa} and bi-gravity~\cite{Crisostomi:2015xia}. An exception to this is the modification to classical general relativity implied by the conformal anomaly, which does not require any screening mechanism, and yet can produce macroscopic effects~\cite{Giannotti:2008cv,Mottola:2010gp,Mottola:2016mpl}.
Members of the second class
include theories like Einstein-dilaton-Gauss-Bonnet gravity~\cite{Gross:1986mw,Metsaev:1987zx,Kanti:1995vq,Yunes:2011we,Charmousis:2014mia,Sotiriou:2014pfa,Yagi:2015oca,Julie:2019sab,Julie:2020vov},
dynamical Chern-Simons gravity~\cite{Jackiw:2003pm, Alexander:2009tp}, and Einstein-\AE{}ther theory~\cite{Jacobson:2000xp,Jacobson:2007veq}. The latter typically include higher curvature modifications to the Einstein-Hilbert action, thus introducing operators of dimension larger than 4, which by dimensional analysis must carry a dimensionful coupling constant or scale. Naively, one may argue that this scale has to be the Planck scale, if the modified theory descends from some quantum gravitational completion of GR. This, however, is not necessarily the case, because the scale could be suppressed, as in the case of the cosmological constant.

\subsection{Graviton mass and propagation speed}

Regardless of the specifics of the theory one considers, there are general properties of the graviton particle (understood as a gauge boson that carries the gravitational interaction) that one may wish to measure or test to ensure our description is as prescribed by Einstein's theory. One such property is the graviton's mass, which according to GR is exactly zero. Theories such as massive gravity~\cite{deRham:2014zqa} and bi-gravity~\cite{Crisostomi:2015xia} predict a non-zero value. In fact, many modified theories created to explain the present-day cosmic acceleration also predict deviations in the propagation of GWs \cite{Cardoso:2002pa, Saltas:2014dha, Lombriser:2015sxa, Lombriser:2016yzn,  Belgacem:2017ihm, Nishizawa:2017nef, Belgacem:2018lbp}, and in the gravitational lensing of GWs~\cite{Congedo:2018wfn,Mukherjee:2019wcg,Mukherjee:2019wfw,Ezquiaga:2020spg}.
GWs have thus the potential to place stringent bounds on the graviton mass, because a non-zero value leads to a modified dispersion relation~\cite{Will:1997bb, Kostelecky:2016kfm}. On very general grounds that rely only on special relativity, a non-zero graviton mass implies that the GW frequency does not just depend on its wave-vector, but rather also on the mass, leading to a compression of the GW train that accumulates with distance travelled~\cite{Will:1997bb}. 

Current GW observations are already placing constraints on the mass of the graviton, but much more can be achieved in the next decade. Current aLIGO/Virgo observations have constrained the graviton mass to be less than $4.7 \times 10^{-23} \; {\rm{eV}}/c^2$~\cite{LIGOScientific:2019fpa}. Constraints on the mass of the graviton, however, can be shown to scale as $[f_{\rm low}/(D_L \rho)]^{1/2}$, where $D_L$ is the luminosity distance, $\rho$ is the SNR, and $f_{\rm low}$ is the lowest frequency detected~\cite{Perkins:2020tra}: this is because the larger the distance, the longer the GW train compression can accumulate for, leading to a stronger constraint. Because of this, in the next few years and then in the next decade, future observations with aLIGO/Virgo/KAGRA/LIGO-India and 3G ground-based detectors can place constraints better than $10^{-25} \; {\rm{eV}}/c^2$ and $10^{-26}$ respectively, while space-borne detectors like LISA can improve these constraints down to $3 \times 10^{-27} \; {\rm{eV}}/c^2$~\cite{Chamberlain:2017fjl,Perkins:2020tra}. These numbers are interesting because if one associates the late-time acceleration of the Universe to a non-zero graviton mass, then the graviton would have to be of the scale of the Hubble constant, $10^{-33}$~eV.  By stacking events from LISA and 3G detectors we may begin to approach this scale, and thus confirm or rule out a non-zero graviton mass as an explanation for the late-time acceleration of the Universe.

Another property of the graviton as a particle that one may wish to probe is its group velocity in the high-energy limit $E \gg m_g$. In Einstein's theory, this group velocity is equal to the speed of light, but in other theories of gravity, this need not be the case~\cite{Baker:2017hug, Ezquiaga:2017ekz, Creminelli2017, Sakstein:2017xjx, Boran:2017rdn, Akrami:2018yjz}. For example, in Einstein-\AE{}ther theory, the graviton travels at a constant group speed that is faster than the speed of light, avoiding causality violations~\cite{Jacobson:2000xp,Jacobson:2007veq}. The measurement of the speed of the graviton, unfortunately, is rather difficult because it requires that we compare the time of arrival of a GW to some other baseline. This is where multi-messenger events shine. If an event produces both GWs and EM waves simultaneously, then one can in principle compare the speed of the GWs to the speed of the EM waves (i.e., the speed of light) by comparing their times of arrival. 

This is exactly what was done with the first aLIGO/Virgo binary neutron star observation, GW170817, which was accompanied by a short gamma-ray burst emitted shortly after merger~\cite{Monitor:2017mdv, Mukherjee:2019wcg, Mukherjee:2019wfw,Baker:2020apq,Ezquiaga:2020dao}. This single observation was sufficient to infer that the speed of the graviton is equal to that of the photon to better than one part in $10^{15}$. Such a measurement had the effect of severely constraining a variety of modified theories of gravity.
Future ground-based observations with aLIGO/Virgo/KAGRA/LIGO-India or with 3G detectors will allow for additional measurements of the speed of the graviton along other lines of sight, and thus allow us to test local position invariance~\cite{Will:2005va,Yunes:2013dva}.

With LISA we may detect supermassive BH binaries at mHz frequencies and measure time delays between the arrivals of photons and gravitons. This will present some advantages. First, the longer timescales of these massive mergers can facilitate triggered EM precursor observations. The inevitable periodic modulations of the EM signal due to Doppler and lensing effects during the inspiral stage arise from the same orbital motion as the GWs, and can be phased in a robust way, without the need to model the astrophysical source in detail~\cite{Haiman:2017szj,Tang:2018rfm}.
The measurements will also provide tighter limits, due to the high SNRs and large horizon distances achievable with LISA.  The frequency dependence of the time delay would further probe Lorentz-violating theories~\cite{Kocsis:2007yu, Haiman:2009te, Mirshekari:2011yq}. 

Some modifications of GR, invoked to explain the present-day cosmic acceleration, predict deviations between the propagation properties of EM radiation and GWs~\cite{Saltas:2014dha,Lombriser:2015sxa,Lombriser:2016yzn,Belgacem:2017ihm,Nishizawa:2017nef,Belgacem:2018lbp,Mastrogiovanni:2020gua}. A multi-messenger data-driven measurement of the running of the effective Planck mass and its redshift dependence is possible by combining three length scales, namely the GW luminosity distance, baryon acoustic oscillations (BAO), and the sound horizon from the CMB~\cite{Mukherjee:2020mha}. Sources detectable at higher redshifts (such as supermassive BH binaries) are most useful to measure the redshift dependence and running of the effective Planck mass. Such measurements may be possible by cross-correlating binary BHs with galaxies~\cite{Mukherjee:2020mha}. GR propagation effects could also be probed using other techniques -- e.g. by using the mass distribution of binary neutron stars~\cite{Finke:2021eio} and BHs~\cite{Leyde:2022orh}.

Weak lensing allows for tests of modified gravity through multiband GW observations~\cite{Holz:1997ic,Congedo:2018wfn,Mukherjee:2019wcg, Mukherjee:2019wfw,Ezquiaga:2020spg,Ezquiaga:2020dao}. Multiband measurements can also be used to test whether constraints on GR deviations are independent of scale, frequency or energy -- for example, reducing to GR at tens of Hz but deviating from GR in the mHz regime~\cite{deRham:2018red}. Propagation tests may also be possible by observing strong lensing of GW signals~\cite{Finke:2021znb}.

\subsection{Polarization of gravitational waves}

\begin{figure}[t]
\centering
\includegraphics[width=0.6\textwidth]{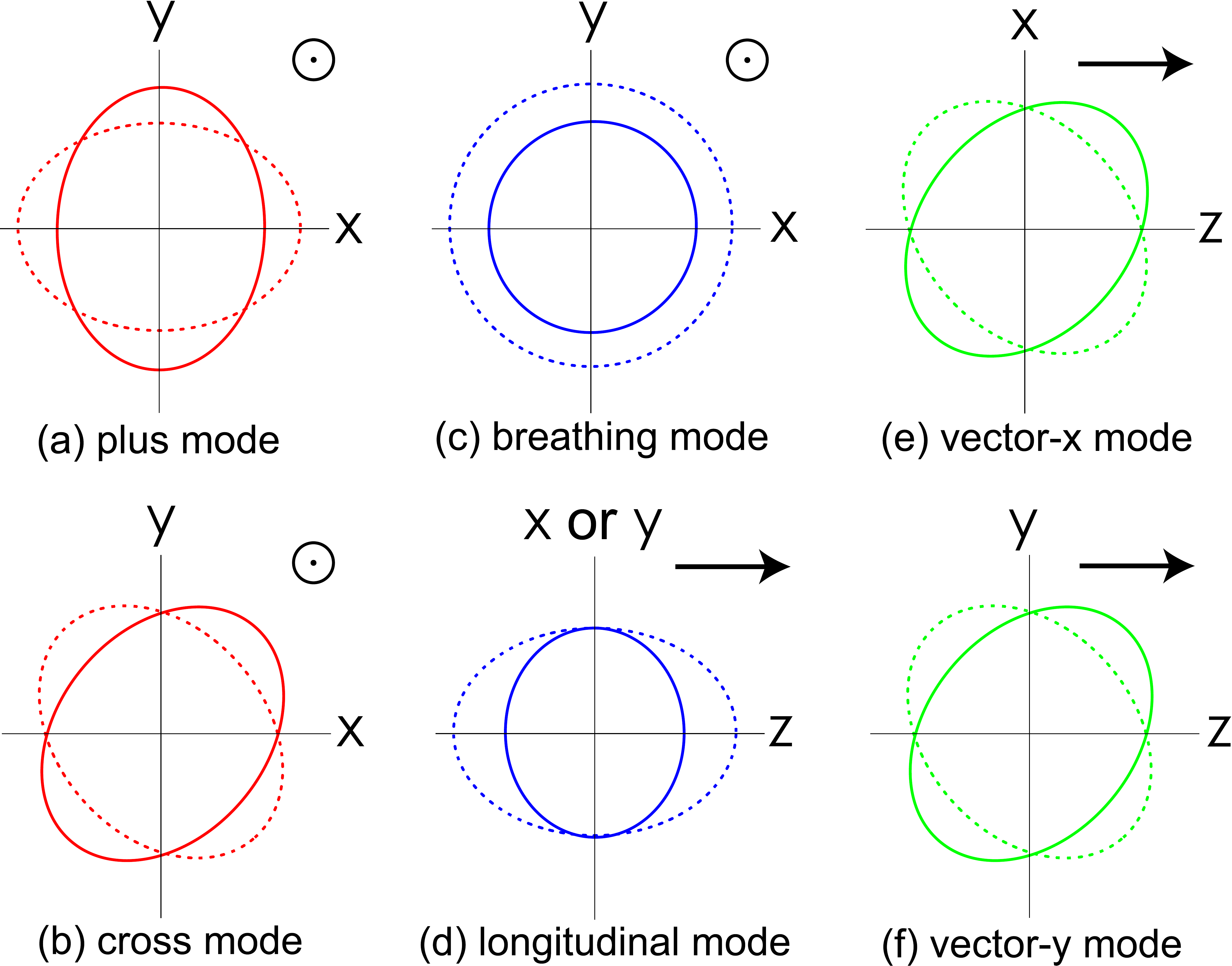}
\caption{The six possible GW polarizations in metric theories of gravity. The solid and dotted lines in each case represent the effect of the GW on a freely falling ring of masses at integer and half-integer multiples of the GW period. GR predicts only plus and cross modes (shown on the left in red), while other metric theories of gravity can predict the existence of more polarizations.  Finding evidence in favor of scalar or vector polarizations would immediately require some modification of classical GR. Reproduced from \citep{Chamberlin:2011ev}.
\label{fig:pols}
}
\end{figure}

A third property of the graviton that one may study is its spin and helicity (projection of the spin along the momentum direction). In GR, the graviton is a spin $s=2$ particle with helicity $+2$ or $-2$, because GWs have only 2 polarizations that can be converted into each other by a rotation of $\pi/(2s) = \pi/4$ radians. The most general GWs, however, can have up to 6 different polarizations~\cite{Eardley:1973br,Eardley:1974nw}: two transverse-traceless tensor modes (like in GR, which are colloquially called the ``plus'' and ``cross'' polarizations), two transverse vector modes, and two scalar modes (colloquially called the breathing mode and the longitudinal mode). 
As shown in Fig.~\ref{fig:pols}, different polarizations will have drastically different effects on matter, as a GW propagates through a detector. But since helicity depends on the direction of propagation of the GW, one must have multiple detectors to allow for enough lines of sight to break degeneracies and extract independent polarization states. Pinning down GW polarization will therefore be a prime target opportunity for a network of future detectors.

Currently, measuring the polarization of GWs is difficult, even though there are 3 GW detectors that can be simultaneously operational. This is because the two aLIGO detectors are not rotated with respect to each other and they are located relatively close to each other (within the United States). This means that effectively there are only 2 linearly-independent data streams (one from the aLIGO instruments and one from Virgo) that can be combined to measure up to 2 independent polarizations. The LVC has therefore only been able to carry out tests to determine whether GWs contain only the 2 tensor polarizations, the 2 vector polarizations or the 2 scalar polarizations~\cite{LIGOScientific:2017ycc,LIGOScientific:2018dkp,Isi:2017fbj,LIGOScientific:2018czr}; tests to constrain the existence of more than the 2 GR tensor polarizations are not yet possible. Once the Japanese KAGRA detector joins the network (in the next few years), the combination of 3 linearly independent data streams will allow for the construction of one null stream: a stream that, if the graviton is a spin-2 particle, will have no signal power in it~\cite{Guersel:1989th,Chatterji:2006nh,Chatziioannou:2012rf}. Once LIGO-India comes online by late this decade or the next, one will be able to construct 2 null streams. In this way, a network of GW detectors will be able to carry out null tests of the spin and helicity content of the graviton. Ground-based detectors will also contribute to the measurement of polarizations by being able to detect signals at lower frequencies ($\sim 1$ Hz), that remain in band long enough for the Earth to rotate and encode the effect of multiple polarizations~\cite{Takeda:2019gwk,Amalberti:2021kzh}.  

Space-based detectors can also carry out tests of the polarization content of GWs. This is because unlike the GWs accessible to current ground-based detectors, those measured by space-based detectors can last months or even years in band. This means that while the GW is being detected, the detector is moving around the Sun, and thus is sampling different lines of sight, which allows for the measurement of multiple polarizations. Similarly, PTAs can also constrain the polarization content of GWs, since (unlike ground-based detectors) they possess many ``interferometer arms'', one per pulsar-Earth system in the array. By correlating the times of arrival of the pulses from multiple pulsars, one can then determine whether their correlation (described by the so-called Hellings-Downs curve in GR) is consistent with a spin-2 particle or not. 

PTAs offer significant advantages over second-generation ground-based interferometers for detecting new polarizations or constraining the polarization content of GWs. Each line of sight to a pulsar can be used to construct an independent projection of the various GW polarizations, and
since PTAs typically observe tens of pulsars, linear combinations of the data can be formed to measure or constrain each of the six polarizations many times over~\citep{Yunes:2013dva}. Additionally, PTAs have an  enhanced response to the longitudinal polarization~\citep{Chamberlin:2011ev}, with inferred constraints on the energy density of longitudinal modes about three orders of magnitude better than constraints for the transverse modes~\citep{Cornish:2017oic}. Using the NANOGrav 9-year data set~\citep{Arzoumanian:2015gjs}, one can set $95\%$ upper limits on the amplitudes of stochastic GW backgrounds from non-GR polarizations at $\Omega_{TT+ST} h^2 < 7.7\times10^{-10}$, $\Omega_{VL}h^2 < 3.5\times10^{-11}$ and $\Omega_{SL} h^2 < 3.2\times10^{-13}$, corresponding to the sum of tensor-transverse and scalar-transverse (breathing) modes, the vector-longitudinal modes, and scalar-longitudinal modes~\citep{Cornish:2017oic}. Very recently, NANOGrav searched their 12.5-year data set for evidence of a GW background with all the spatial correlations allowed by general metric theories of gravity~\cite{NANOGrav:2021ini}. They found no substantial evidence in favor of the existence of non-Einsteinian correlations, and placed upper limits on the amplitudes of the stochastic background produced by eight different families of metric theories of gravity.

\subsection{Symmetries of the gravitational sector}

Aside from probing the particle properties of the graviton, GWs also allow us to investigate the fundamental symmetries at play in the gravitational sector. This search for fundamental symmetries resembles the development of the Standard Model of particle physics, which was guided by results from particle colliders. One can similarly ask if the foundations of Einstein's gravity theory rest on solid experimental ground. A symmetry one may probe is gravitational Lorentz invariance. Several ultraviolet completions of GR predict that Lorentz symmetry may be spontaneously broken at some small scale in the gravitational sector. In some theories, this has been found to be necessary for the theory to be power-counting renormalizable~\cite{Horava:2009uw,Barvinsky:2015kil}, as in the case of Ho\v{r}ava and khronometric gravity~\cite{Blas:2009qj,Jacobson:2010mx}. The most general theory that breaks Lorentz symmetry while retaining second-order field equations is Einstein-\AE{}ther theory~\cite{Jacobson:2000xp,Jacobson:2007veq}, which is why attention on constraining this model has spiked in recent years.     

The multi-messenger GW170817 event described earlier has already placed constraints on gravitational Lorentz violation. This is because the two tensor GW polarizations in this theory propagate faster than the speed of light. This constraint, however, limits only 1 combination of the 4 coupling constants in Einstein-\AE{}ther theory, and there is still a large coupling phase space~\cite{Barausse:2019yuk} that allows for spontaneous Lorentz symmetry breaking (even after accounting for Solar System~\cite{Will:2014kxa,Foster:2005dk,Blas:2011zd,Bonetti:2015oda,Muller:2005sr}, binary pulsar~\cite{Yagi:2013qpa,Yagi:2013ava,Gupta:2021vdj} and cosmological constraints~\cite{Carroll:2004ai,Audren:2014hza}). GW observations of neutron stars and BHs may allow further constraints~\cite{Hansen:2014ewa,Zhang:2019iim}, and tests of gravitational Lorentz violation are today limited by the lack of theoretical models to compare against the data~\cite{Hansen:2014ewa,Zhang:2019iim}. Once further modeling is completed, tests with current and future detectors will place more stringent bounds. 

Another fundamental symmetry one may wish to probe is parity. In Einstein's theory, parity is preserved under a reversal of the spatial triad. In the standard model of particle physics, however, CP symmetry
is broken, as famously shown from the decay of neutral kaons~\cite{Christenson:1964fg}, which earned Cronin and Val Fitch the Nobel Prize in Physics in 1980. The question then arises as to whether parity could be broken in the gravitational sector, an effect that can be modeled in dynamical Chern-Simons gravity~\cite{Jackiw:2003pm,Alexander:2009tp}. This effective theory of gravity modifies the Einstein-Hilbert action through a pseudo-scalar field that couples to the Pontryagin topological invariant. As such, it induces a variety of modifications in the generation of GWs and also in their propagation. Parity violation is most clearly manifest in the latter, where left- and right-polarized GWs obey different propagation equations~\cite{Alexander:2007kv,Yunes:2010yf}. The detection of an imbalance in left- and right-polarized modes would then be a signal of parity violation~\cite{Yunes:2010yf,Okounkova:2021xjv}. 

GW detectors are currently searching for such effects and attempting to constrain them, but this has proven to be challenging. The excess or deficit of left- versus right-polarized GWs can only be established when enough sources are detected to construct robust population statistics~\cite{Okounkova:2021xjv}. The LVC has carried out the first tests of this type, but constraints will become much more robust when 3G detectors observe thousands of sources. A multi-messenger event could also be used to constrain parity violation, because source localization and the detection of (say) short-gamma ray bursts would break degeneracies with the parity-violating modifications~\cite{Yunes:2010yf}. The GW170817 event was not sufficiently clean to carry out this test, so we will have to wait for future multi-messenger events with 3G detectors. The generation of GWs is also modified by parity-violating interactions, most notably activating a pseudo-scalar wave that carries energy away from a binary BH, forcing it to inspiral faster than in GR~\cite{Yagi:2011xp,Yagi:2012vf,Yunes:2016jcc}. This effect, however, is degenerate with the spin angular momenta of the BHs in the binary, and thus, the test has so far been ineffective with current data. As ground-based detectors are improved in the next few years, it is likely that this test will be performed, and we will be able to constrain the degree to which parity could be violated in the gravitational interaction~\cite{Alexander:2017jmt}.  

\section{Black hole horizons, quantum gravity, and the information paradox}
\label{sec:BHs}

At the most microscopic scales probed by experiments so far, the principles of
quantum field theory (QFT) hold.
Yet Einstein's classical GR remains unreconciled with quantum theory, and tests 
of GR are still based on an essentially classical description of matter and energy.
This tension between quantum matter and classical gravity comes to the fore in the
puzzles and paradoxes of BHs.

Classical BHs are vacuum solutions of Einstein's equations, in which all details 
of the quantum matter that gave rise to them are subsumed into an interior 
spacetime singularity of infinite pressure and density. According to the singularity
theorems, as long as a closed trapped surface forms and the collapsing matter
satisfies certain energy conditions, a BH singularity is 
unavoidable~\cite{Penrose:1964wq,Penrose:1969pc}. In the case of rotating BHs,
analytic extension of the exterior Kerr solution through the horizon 
leads not only to singularities but also to closed timelike curves in 
their interior, violating causality at even macroscopic scales. 

For these reasons it is virtually certain that the purely classical description 
breaks down in the BH interior, or possibly even at the horizon boundary of this
region. In classical GR the interior region is defined by the existence of a trapped
surface and of an event horizon causally disconnecting it from observers at future
infinity. Yet the presumption that gravitational collapse of very massive stars
leads inevitably to a BH horizon and interior singularities is based upon an
essentially classical view of the collapsing matter, and the assumption of no
energy-momentum sources at the horizon. Both of these assumptions can be questioned,
since Standard Model matter is certainly quantum in nature, and quantum matter may
not satisfy the assumed classical energy conditions.

When quantum effects are considered, additional problems appear at the macroscopic 
horizon scale. Since a BH horizon is a marginally trapped surface from which no 
matter or information can escape, at least classically, the entropy of matter 
falling into a BH would seem to have vanished from the external Universe. On the
other hand, if the BH horizon itself contributes an entropy equal to one quarter 
of its area in Planck units -- as Bekenstein suggested, and as seemingly implied 
by the Hawking effect~\cite{Bekenstein:1973ur,Hawking:1974rv,Hartle:1976tp} -- it 
is not clear how this enormous entropy can arise from a counting of microscopic
states of a surface where nothing special is supposed to happen locally in 
classical GR. BH entropy also gives rise to a potential conflict with quantum
unitary evolution, a cornerstone upon which quantum physics itself is founded. 
As a result the {\it ``BH information paradox"} has stimulated theoretical research 
and debate for over four decades
~\cite{Unruh:2017uaw,Mathur:2005zp,Giddings:2017mym}. Quantum physics however 
allows for nonlocal effects at the horizon scale~\cite{Mottola:2010gp}, 
which may lead to a physical surface and an entirely different interior (that of 
a gravitational vacuum condensate) which is free of both spacetime singularities 
and any information paradox~\cite{Mazur:2001fv,Mazur:2004fk, Mazur:2015kia}. 

With the increase in GW and multi-messenger data anticipated in this decade, tests 
of this and other hypotheses for solving the BH paradoxes arising from the tensions
between QFT and classical GR will become possible for the first time. We
are therefore on the threshold of transforming BH physics from a theoretical
conundrum to a subject of observational science, with potentially far-reaching
implications for the foundations of physics, including the quantum nature of gravity.

\subsection{Gravitational-wave and electromagnetic tests of black hole spacetimes}

One of the outstanding observational quests concerns testing GR in the strong-field regime. Many of the conceptual issues listed above can be better understood by placing experimental constraints on BH spacetimes and on the existence of horizons. Uniqueness results in vacuum GR imply that isolated BHs are fully described by only three parameters (mass, spin, and possibly electric charge), making them the simplest macroscopic objects in the Universe~\cite{Chrusciel:2012jk}. In addition, BHs posses the tantalizing property that they don't ``polarize'' under the influence of a companion~\cite{Binnington:2009bb,Damour:2009vw,Cardoso:2017cfl,LeTiec:2020bos,Chia:2020yla}.
The simplicity of BHs, whether isolated or in binaries, implies that they are ideal laboratories to probe classical GR.

Dynamical tests of gravity in its strong-field, highly dynamical and nonlinear regime can be done to exquisite precision with GWs. Consistency with GR requires the {\it full merger signal} to be consistent with the theory. This ultimate test of GR requires large SNRs as well as accurate GR calculations of the waveforms emitted by binary systems throughout the merger, so that systematic modeling errors do not affect the tests.

Tests of generic theories of gravity and BH spacetimes beyond GR would require dynamical evolutions. This is a difficult problem because of the complexity of the equations of motion and, more fundamentally, because several modifications of GR lack a well-posed initial-value problem~\cite{Brito:2014ifa,Papallo:2017qvl,Bernard:2019fjb,Julie:2020vov}. Therefore many proposed tests rely either on slow-motion parametrizations of the early (inspiral) phase~\cite{Yunes:2009ke}, or on late-time expansion of the GW signal based on BH perturbation theory~\cite{Berti:2005ys}. 

In particular, the uniqueness properties of BHs in GR imply that such objects have a simple multipolar structure, which specifies uniquely the expansion of their gravitational field. The relativistic multipole moments of a stationary, asymptotically flat spacetime are defined in terms of two sets of
quantities evaluated
at infinity: the {\it mass multipole moments} $M_\ell$ and the {\it current multipole moments} $S_\ell$. In the Newtonian limit, the mass multipole moments reduce to the moments in Newtonian
theory~\cite{Geroch:1970cd,Hansen:1974zz,Cardoso:2016ryw}. For the Kerr BH spacetime with mass $M$ and angular momentum per unit mass $a\leq M$,
\begin{eqnarray}
M_{2 \ell}&=&(-1)^{\ell}Ma^{2 \ell}\,,\nonumber\\
S_{2 \ell+1}&=&(-1)^{\ell}Ma^{2 \ell+1}\,.\label{Kerrmoments}
\end{eqnarray}
In other words, all multipole moments are fixed in terms of the mass and spin alone, a manifestation of the ``no-hair'' properties of BHs in GR.

One of the most natural ways to test the spacetime metric of a BH is to study -- through astrophysical or GW observations -- the motion of stellar objects in its surroundings. If a stellar-mass BH or neutron star orbits a supermassive object, the inspiral process (driven by GW emission) will be sensitive to the entire multipolar structure of the central object~\cite{Ryan:1995wh, Krishnendu:2017shb, Krishnendu:2018nqa} (in addition to there being an horizon or not). A comparison with Eq.~(\ref{Kerrmoments}) then provides a theory-agnostic test of GR.

The measurement of the different multipole moments can also be done in the EM band. The multipole moments of BH candidates can be measured from EM observations of stars on tight orbits around supermassive BHs, and in particular Sgr A*, the compact object at the center of the Galaxy~\cite{Will:2007pp,Merritt:2009ex,Psaltis:2015uza,Christian:2015smg}.
Such tests are complicated by several additional factors (related to the fact that such orbits have typical radii much larger than those involved in GW observations), but progress in instrumentation makes them attractive
possibilities in the near-future.
The key idea is to measure the pericenter and orbital plane precession of stars orbiting a massive BH, on tight and eccentric enough orbits. The precession depends on the mass, spin and quadrupole moment of the central object, and therefore the measurement can be inverted to estimate each of these quantities and to test the BH nature of the object. 

The observation of a {\it single} pulsar in orbit around a very compact object may also allow for tests of the Kerr hypothesis~\cite{Wex:1998wt} using pulsar timing data (see e.g.~\cite{Liu:2011ae,Psaltis:2015uza} for a description of the general strategy and sources of error).
The mass of the central object and the inclination of the orbital plane can be determined from the precession of the periastron or via Shapiro time delay. 
The Lense-Thirring precession, along with a measurement of the periastron precession, the projected semi-major axis and their time derivatives, allow for the determination of all three spin components. Finally, Roemer delay can be used to estimate the quadrupole moment~\cite{Liu:2011ae}.
Tests of the Kerr hypothesis using single pulsars require high-eccentricity pulsars and sub-year orbital periods, but they are a promising prospect for the near future~\cite{Liu:2011ae,Psaltis:2015uza}.

When matter moves {\em very close} to a BH (or any compact object) a multipolar decomposition of the gravitational field is not particularly useful, since all (or a substantial number of) multipoles contribute to the gravitational potential and to the motion of matter. However, the spacetime around compact objects possesses unique features -- such as innermost stable circular orbits and light rings (unstable null geodesics)~\cite{Bardeen:1972fi} -- that might be used as smoking guns of the BH nature of the object and of GR. In particular, null (photon) geodesics carry information about the effective size of BHs, since in essence any particle or light ray penetrating the light ring will never reach asymptotic observers: this is sometimes referred to as the BH ``shadow''~\cite{1975ApJ...202..788C,1979AA....75..228L,Falcke:1999pj}.
The exact shape and appearance of BHs depends on the source illuminating them and on the BH rotation rate, which determines how close to the horizon the co- and counter-rotating light rings are, how tightly the accretion disk can bind to the BH, and the magnitude of the gravitational and Doppler shifts from the disk.
Observations of BH shadows became possible with the advent of powerful instruments such as the Event Horizon Telescope~\cite{EventHorizonTelescope:2019dse}.
The main obstacles to performing tests of GR in this way are
(i) the large number of parameters that describe the shadow (including the inclination angle of the object, the mass and angular momentum of the BH, and the details of the accretion mechanism);
(ii) the lack of a robust parametrization of strong-field deviations from the Kerr geometry. 

Multimessenger observations hold the promise of providing additional tests. If an EM light-curve can be obtained for a massive LISA BH binary, prior to the merger, over at least several orbital cycles, then there is a significant probability of seeing unmistakable, periodic self-lensing flares whenever the two BHs are aligned along the line of sight, within the Einstein radius of the foreground (lens) BH~\cite{Haiman:2017szj,DOrazio:2017ssb}. Furthermore, for compact binaries in the LISA band, the BH horizon and/or photon rings should imprint characteristic features (``dips'') near the peaks of the flares in this light-curve~\cite{Davelaar:2021eoi,Davelaar:2021gxx}. These dips carry information analogous to the size and shape of the spatially resolved ``shadow'' measured by the Event Horizon Telescope, but would be seen ``directly'' in the light-curves, and hence do not require high angular resolution.

All of the above tests are dependent on the motion of matter on an otherwise {\em fixed} BH background. When a massive companion is present, tidal effects must be taken into account, but BH binaries are very special: their so-called tidal Love numbers and tidally induced multipole moments vanish identically~\cite{Binnington:2009bb,Damour:2009vw,Cardoso:2017cfl,LeTiec:2020bos,Chia:2020yla,Poisson:2021yau}. Therefore, accurate tracking of the GW phase allows for constraints on the tidal properties of the inspiraling objects and for tests of their BH nature~\cite{Maselli:2017cmm}.

\subsection{Tests of black hole dynamics and black hole spectroscopy}

The dynamical content of the underlying theory of gravity can be probed in violent, dynamical situations giving rise to strong bursts of GW emission. After the violent merger of two compact objects leading to BH formation, GR predicts the formation of a Kerr BH, so that the spacetime is described by only two parameters. The relaxation to this state is described by a set of exponentially damped sinusoids (``ringdown'') whose frequencies and damping times depend only on the mass and spin~\cite{Kokkotas:1999bd,Berti:2009kk}. Since GW observations provide a measurement of frequencies and damping times, the ``ground state'' QNM allows us to infer the mass and spin. Any measurement of additional QNM frequencies (``excited states'') can then be used as a null test of the Kerr nature of the remnant. 

The idea of treating BHs as ``gravitational atoms'', thus viewing their QNM spectrum as a unique fingerprint of spacetime dynamics (in analogy with atomic spectra), is usually referred to as ``BH spectroscopy''~\cite{Dreyer:2003bv,Berti:2005ys,Berti:2007zu,Gossan:2011ha, Meidam:2014jpa}.  The seeds of this idea were planted in the 1970s (see e.g.~\cite{Berti:2009kk} for a detailed chronology).
Chandrasekhar and Detweiler developed various methods to compute the QNM spectrum, identifying and overcoming some of the main numerical challenges (see e.g.~\cite{Chandrasekhar:1975zza}). In particular, Detweiler concluded the first systematic calculation of the Kerr QNM spectrum~\cite{Detweiler:1980gk} with a prescient statement: {\em ``After the advent of gravitational wave astronomy, the observation of [the BH’s] resonant frequencies might finally provide direct evidence of BHs with the same certainty as, say, the 21 cm line identifies interstellar hydrogen.”}

Early estimates~\cite{Berti:2005ys,Berti:2007zu} showed that the detection and extraction of information from ringdown signals requires events whose SNR in the ringdown {\em alone} is larger than those achievable now (for example, the first GW detection (GW150914) had a combined SNR of $24$, with an SNR$\sim 7$ in the ringdown phase~\cite{TheLIGOScientific:2016wfe,TheLIGOScientific:2016src}). There are claims that overtones have been detected in GW150914~\cite{Isi:2019aib} and higher modes have been measured in GW190521~\cite{Capano:2021etf}, but the detection of modes other than the fundamental is debatable at current SNRs~\cite{Bustillo:2020buq,LIGOScientific:2020tif,Cotesta:2022pci}. 
Any deviation from the QNM spectrum of classical GR would indicate substructure of 
BH ``atoms'' inconsistent with the standard picture. In particular, a non-singular horizonless object would lead to different boundary conditions than the classical theory and departures from the BH QNM spectrum.
In any case, conclusive tests should be achievable once Advanced LIGO and Virgo reach design sensitivity, and certainly with 3G detectors (Cosmic Explorer or the Einstein Telescope) or with space-based detectors such as LISA~\cite{Berti:2016lat}. If the frequencies turn out to be compatible with the predictions of GR, parametrized formalisms can be used to constrain theories of gravity that would predict different spectra~\cite{Cardoso:2019mqo,McManus:2019ulj,Maselli:2019mjd,Carullo:2021dui}.

\subsection{Testing the existence of horizons}

The existence and properties of horizons can be inferred and quantified with a variety of observations~\cite{Cardoso:2019rvt}. 
It is believed that accreting horizonless objects would reach thermal equilibrium with the environment rather quickly, whereas accreting supermassive BHs do not: the luminosity contrast between the central accreting object and its accretion disk imposes stringent constraints on the location and property of a putative surface~\cite{Broderick:2009ph,Cardoso:2019rvt}. 
However constraints based on accretion models are model dependent, and have also been questioned~\cite{Carballo-Rubio:2018jzw}. They still leave open the possibility
of a surface close to the would-be event horizon, as predicted in thin shell gravastar models~\cite{Mazur:2004fk,Mazur:2015kia,Beltracchi:2021zkt,Beltracchi:2021lez}. The planned EHT and future surveys of tidal disruption events will improve current constraints on the location of a hypothetical surface by two orders of magnitude. 

The EM observations above are done essentially in a fixed-background context, in which the BH spacetime is an arena where photons propagate. One can also consider situations probing both the background {\it and} the field equations. A stellar-mass BH or a neutron star orbiting a supermassive BH will slowly inspiral due to emission of GWs, ``sweeping'' the near horizon geometry and being sensitive to tiny near-horizon changes, such as tidal deformability or tidal heating, or to non-perturbative phenomena like resonances of the central object~\cite{Cardoso:2017cfl,Maselli:2017cmm,Cardoso:2019nis,Maggio:2021uge,Fang:2021iyf}.
Accurate tracking of the GW phase by the future space-based detector LISA may constrain the location of a putative surface to Planckian levels~\cite{Cardoso:2019rvt}. 

The absence of an horizon can also lead to smoking-gun effects in the GW signal. An ultracompact, horizonless vacuum object sufficiently close to the Kerr geometry outside the horizon behaves as a cavity for impinging GWs, which end up being trapped between the object's interior and its light ring~\cite{Cardoso:2016rao,Cardoso:2016oxy,Cardoso:2019rvt}. Thus, perturbations of such objects, and possibly mergers as well, lead to a GW signal which is -- by causality principles -- similar to that emitted by BHs on sufficiently small timescales. However, at late times the signal trapped in the ``cavity'' leaks away as a series of ``echoes'' of the original burst, which may carry a significant amount of energy. LIGO/Virgo observations have so far shown no evidence for such echoes~\cite{LIGOScientific:2020tif,LIGOScientific:2021sio}. The absence of such structure in future observations by LIGO and LISA will allow the exclusion -- or detection -- of any significant structure a Planckian distance away from the Schwarzschild radius, with important implications for fundamental physics~\cite{Cardoso:2019rvt}. 

Setting stringent constraints on the nature of compact objects -- in particular quantifying the existence of horizons in the Universe -- requires advanced detectors. It is also a challenging task from the modelling and computational point of view, as one needs: (i) a physically motivated, well-posed theory solving -- at least partially -- the conceptual problems of GR; (ii) the existence in such theories of ultracompact objects which arise naturally as the end-state of gravitational collapse; and (iii) the solution of the relevant partial differential equations describing the mergers of such objects. There is pressing need for progress on all of these fronts to confront the increasingly precise data expected from a wide variety of new experimental facilities. 

\section{Gravitational signatures of dark matter}
\label{sec:DM}

Dark matter (DM) is the dominant part of the matter density of the Universe. Together with the fate of gravitational singularities, the nature of DM has been for decades -- and remains today -- one of the outstanding issues in physics. Many Snowmass White Papers discuss open problems in this field, which is too large and complex to summarize here. Our main focus will be on possible gravitational signatures of DM.

\subsection{Dark compact objects}

Some of the DM may have clustered gravitationally in the early Universe, forming compact dark objects. These structures may cause a transient magnification of light from distant stars via microlensing, which remains one of the most powerful techniques to constrain compact dark objects in a wide range of masses~\cite{1986ApJ...304....1P}.

DM clumps near (or within) the Earth can alter the planet's tidal field -- which is well monitored for decades and therefore well constrained -- or cause sudden accelerations, leading to interesting constraints on asteroid-like clumps~\cite{Seto:2007kj,Namigata:2022vry,Kashiyama:2018gsh}. Albeit small, the interaction cross section of DM with Standard Model fields can lead to the deposition of small DM cores at the center of stars~\cite{Press:1985ug}, with capture rates that can be enhanced by the large density of white dwarfs and neutron stars. For fermionic fields, the accumulation of DM could eventually lead to cores more massive than the Chandrasekhar limit, collapse of the DM core to a BH, and eventually to the disruption of the star by accretion onto the newly formed BH~\cite{Goldman:1989nd}. For bosonic DM, this fate may be eluded via gravitational cooling~\cite{Brito:2015yga}.

Another possibility is that standard cold dark matter (CDM) models could produce small-scale clumps. A CDM clump moving near the Earth or a pulsar produces an acceleration that could be measurable in PTA data, providing an opportunity to test the CDM paradigm~\citep{Siegel:2007fz,Kashiyama:2018gsh}. 

The possibility of compact objects harboring DM cores is intriguing. If these cores are sufficiently massive the star is effectively described by a different equation of state, and its properties change. The coalescence of DM stars will differ from the prediction of standard GR, leading to peculiar signatures in the GW signal close to merger~\cite{Ellis:2017jgp,Bezares:2019jcb}.
In fact, DM clumps can also form in isolation and bind to compact stars in their vicinity. Compact DM cores orbiting neutron stars (either in their exterior or in their interior) may give rise to detectable signals in our Galaxy~\cite{Horowitz:2019aim}.

The general GW signatures of the coalescence of DM clumps or ``blobs'' have been explored by various authors (see e.g.~\cite{Giudice:2016zpa,Diamond:2021dth}), but precise calculations of the signal from the coalescence of two DM clumps require an underlying theory with a well-posed initial value problem. One example are compact configurations made of self-gravitating scalar fields, also known as boson stars~\cite{Palenzuela:2017kcg,Cardoso:2017cfl,Sennett:2017etc,Bustillo:2020syj}.

A well-studied class of DM cores consists of primordial BHs (PBHs), which have been predicted as a generic outcome of density perturbations in the early Universe~\cite{1967SvA....10..602Z,1971MNRAS.152...75H,1975ApJ...201....1C, 1980PhLB...97..383K, 1985MNRAS.215..575K,Carr:2005zd, Clesse:2016vqa, Sasaki:2018dmp, Sasaki:2016jop, Raidal:2017mfl, Raidal:2018bbj, Vaskonen:2019jpv, Gow:2019pok,Jedamzik:2020ypm,Jedamzik:2020omx,DeLuca:2020jug, Atal:2020igj, DeLuca:2020qqa,Clesse:2020ghq}. PBHs can exist over a wide range of masses ranging from around $10^{-16}\,M_\odot$ to $10^{10}\,M_\odot$, and they may form a significant fraction of the DM~\cite{Bird:2016dcv,Ali-Haimoud:2017rtz, Raidal:2017mfl, Raidal:2018bbj, Vaskonen:2019jpv, Atal:2020igj, DeLuca:2020qqa, Wong:2020yig, Franciolini:2021tla}. GWs may allow us to detect PBHs over a wide range of masses, being complementary to other proposed probes, such as gravitational lensing, dynamical friction, and CMB distortions. A firm detection requires the ability to distinguish PBHs from BHs of astrophysical origin. PBHs can be distinguished on the basis of their source properties such as mass, and eccentricity spin, the redshift evolution of BBH merger rates, and their spatial distribution. 
The GW signal can distinguish between BHs of astrophysical origin and PBHs using either resolved events~\cite{Hall:2020daa,Wong:2020yig,DeLuca:2021wjr,Hutsi:2020sol,Franciolini:2021tla} or the stochastic GW background~\cite{Mandic:2016lcn,Mukherjee:2021ags,Mukherjee:2021itf}.
The detection of sub-solar BHs would be a smoking gun of the existence of PBHs. BHs of both sub-solar and higher masses may be observed at high redshift, either through their stochastic GW background \cite{Mukherjee:2021ags, Mukherjee:2021itf} or as individual events observed with 3G GW detectors. The detection of BBHs of any mass at high redshift, where star formation should be rare, can also provide a smoking gun signature of PBHs~\cite{Raidal:2017mfl, Raidal:2018bbj, Vaskonen:2019jpv, Atal:2020igj,Mukherjee:2021ags, Mukherjee:2021itf, Ng:2021sqn,Franciolini:2021xbq}. 
Therefore 3G detectors could give us conclusive evidence of whether PBHs form a significant fraction of DM in a wide range of masses. LISA and PTAs may observe stochastic GW signals from PBHs, yielding independent constraints on their existence~\cite{Kohri:2018awv, Espinosa:2018eve, Wang:2019kaf}. The detection of PBHs would not only shed light on DM candidates, but also on the physics of the early Universe.

\subsection{Dilute DM distributions}
One of the most solid experimental pillars of modern physics is the equivalence principle, which ensures that all forms of matter couple universally to gravity. Even if DM does not form compact object, dilute DM configurations must still interact gravitationally. Dense DM spikes can then develop in the vicinity of isolated compact bodies such as BHs~\cite{Gondolo:1999ef,Sadeghian:2013laa}. Massive BHs are expected to be present at the center of many galaxies. In these environments the DM density should therefore be substantially higher than in the Solar System. Compact objects (BHs or neutron stars) moving in these dense DM environments will be subject to accretion and dynamical friction, leading to small changes in their dynamics that require a detailed understanding of the physics involved in these processes. Preliminary studies indicate that DM-induced changes in the GW phase of compact objects could be detectable by 3G interferometers~\cite{Barausse:2014tra,Cardoso:2019rou,Kavanagh:2020cfn,Annulli:2020lyc,Annulli:2020ilw,Traykova:2021dua,Vicente:2022ivh}.

If DM has a very large Compton wavelength, as in the case of ``fuzzy'' ultralight DM fields of mass $10^{-23} - 10^{-22}$ eV, it may give rise to small pressure oscillations at low frequency (e.g. of the order of nanoHz), that could affect the motion of stars and binary systems~\cite{Khmelnitsky:2013lxt,Porayko:2018sfa}. These minute changes can be tracked with PTA experiments.
In fact, these oscillations can affect the GW detectors themselves: the direct couplings to the beam splitter of GW detectors can be used to set stringent constraints on the abundance and coupling strength of DM~\cite{Vermeulen:2021epa,Pierce:2018xmy}.

\subsection{Nonperturbative effects: ultralight bosonic fields}
The simplest possibility for new matter sectors are bosonic or fermionic degrees of freedom minimally coupled to gravity. These fields could form all or part of the DM. Their scale is set by their mass $\mu$, which could range from cosmological scales to very heavy particles~\cite{Dine:1982ah,Preskill:1982cy,Arvanitaki:2009fg}. Bosonic fields with Compton wavelength comparable to the Schwarzschild radius of astrophysical BHs of mass $M$, i.e.  $GM\mu/(c\hbar)\sim 1$, can trigger a new fascinating phenomenon caused by the existence of {\it ergoregions} around spinning BHs~\cite{Penrose:1969pc,Brito:2015oca}. Spinning BHs can spontaneously transfer their rotational energy to a boson ``condensate'' or ``cloud'' co-rotating with the BH and carrying a significant fraction of its angular momentum. The bosonic cloud is a classical object of size much larger than the BH itself, and it can contain up to $10\%$ of its mass~\cite{Brito:2015oca}. The BH/cloud system is similar to a huge gravitational ``lighthouse'' which extracts energy from the BH by emitting a nearly monochromatic GW signal.

Proposed ways to rule out or constrain light bosons as DM candidates include~\cite{Brito:2015oca}:

\begin{itemize}
    \item[(i)] Monitoring the spin and mass distribution of astrophysical BHs. Measurements of highly spinning BHs will immediately rule out fields with Compton wavelengths comparable to the horizon radius, as these BHs should have been spun down on relatively short timescales.

    \item[(ii)] Direct searches for the resolvable or stochastic  monochromatic GW signals produced by the boson cloud~\cite{Arvanitaki:2010sy,Arvanitaki:2014wva,Brito:2014wla,Arvanitaki:2016qwi,Brito:2017wnc,Brito:2017zvb,Ghosh:2018gaw}, which are now routinely carried out by the LIGO/Virgo collaboration~\cite{LIGOScientific:2021jlr}.

    \item[(iii)] Searches for EM emission from BH/boson cloud systems. Axion-like particles have been proposed in many theoretical scenarios, including variations of the original solution to the strong CP problem of QCD. Self-interactions and couplings with Standard Model fields can lead to periodic bursts of light, ``bosenovas'' and other interesting phenomenology~\cite{Yoshino:2012kn,Ikeda:2018nhb}. In addition, axion-like particles should couple to photons and produce preferentially polarized light~\cite{Chen:2019fsq}.

    \item[(iv)] Observations of peculiar stellar distributions around massive BHs. The nonaxisymmetric boson cloud can cause a periodic forcing of other orbiting bodies, possibly leading to Lindblad or corotation resonances  where stars can cluster~\cite{Ferreira:2017pth,Boskovic:2018rub}.
\end{itemize}

These are only some of the possible strategies.  Superradiance does not require any "seed" boson abundance: any vacuum fluctuations will lead to energy extraction and grow exponentially in time. In this sense BHs are natural particle detectors, complementary to terrestrial colliders~\cite{Brito:2015oca,Barack:2018yly}.  It is important to remark the complementary role of the different GW and EM instruments necessary to probe the large range of mass/energy scales: astrophysical BHs span about ten orders of magnitude in mass, thus allowing us to constrain ultralight bosonic fields across ten orders of magnitude in mass (or energy).

Most of our discussion focused on a neutral DM environment and on gravitational dynamics. Another possibility is that beyond the Standard Model fermions may carry a fractional electric charge, or be charged under a hidden $U(1)$ symmetry~\cite{DeRujula:1989fe,Perl:1997nd}. Modified theories of gravity can also lead to compact stars or BHs carrying nonzero scalar charges~\cite{Damour:1993hw,Doneva:2017bvd,Silva:2017uqg}.  In all of these theoretical scenarios, BHs and compact stars can carry non-negligible charges that would lead to different inspiral and merger signals~\cite{Zilhao:2012gp,Cardoso:2016olt,Alexander:2018qzg,Kopp:2018jom,Dror:2019uea,Bozzola:2020mjx,Maselli:2021men}: GW observations can be used to reveal or constrain these charges and the underlying theories.

\section{Mapping the expansion history of the Universe using multi-messenger observation}
\label{sec:MMA}

\subsection{Hubble constant measurement}

Recently, two main probes of the Hubble constant $H_0$ have come into significant tension: the latest measurement from the Cepheids and SNe Ia distance ladder are discrepant at the $5\sigma$ level with \emph{Planck} observations of the CMB~\cite{Abbott:2017smn,Planck:2018vyg,Wong:2019kwg,Verde:2019ivm,Freedman:2019jwv,Riess:2019cxk,Blakeslee:2021rqi,Riess:2021jrx,Freedman:2021ahq,DiValentino:2021izs,Shah:2021onj}. As an independent, new probe of $H_0$, GW standard sirens could help us to clarify the origin of this tension~\cite{Schutz:1986gp,Holz:2005df,Abbott:2017xzu, Abbott:2019yzh,Virgo:2021bbr}. Standard sirens are ideal for this purpose as they are ``self-calibrating'' distance indicators, i.e., unlike SNe, they do not require a distance ladder: they are calibrated directly by the theory of general relativity. While the GW signal can provide a direct luminosity distance measurement, the mass-redshift degeneracy requires external information to derive a redshift, and thus to probe $H_0$ and other cosmological parameters through the distance-redshift relation. Here we focus on multi-messenger standard sirens that require the combination of EM observations with GW data. The GW sources can be broadly classified in two classes: (i) bright standard sirens, from which we expect EM counterparts; and (ii) dark standard sirens, from which we do not expect EM counterparts.

\noindent {\bf \em Bright standard sirens:} EM counterparts from GW sources are expected from compact objects such binary neutron stars, neutron star-black hole binaries, and BH binaries surrounded by baryonic matter~\cite{Schutz:1986gp,Holz:2005df,Dalal:2006qt,Coulter:2017wya,LIGOScientific:2017ync}. The LIGO/Virgo detection of the binary neutron star event GW170817, jointly with multiband EM observations ranging from gamma-ray to radio~\cite{Abbott:2017xzu,TheLIGOScientific:2017qsa}, ushered in the era of multimessenger astronomy with bright standard sirens.  The constraining power of standard sirens on $H_0$ is most prominent for low-redshift binary neutron stars ($z\lesssim 0.1$) observed by the LIGO/Virgo/KAGRA detector network with an identified host galaxy~\cite{Schutz:1986gp,Holz:2005df,Dalal:2006qt,MacLeod:2007jd,Nissanke:2009kt,Abbott:2017xzu,Chen:2017rfc,Feeney:2018mkj}. Similarly, neutron star-BH binaries should be detected at low redshift, and EM counterparts are possible (depending on the source properties).

In addition, LISA should be able to measure the luminosity distance to inspiraling supermassive BH binaries up to redshift $z\sim 10$, prior to the onset of dark energy domination and well beyond what SNe can probe~\cite{Cutler:2009qv,Tamanini:2016zlh}. LISA measurements in coordination with EM observatories can yield few-percent constraints of the value of the Hubble constant. EM emission is possible because massive BH mergers take place in galactic nuclei, where large amounts of gas can surround the merging binary.  Recent numerical simulations suggest that a massive BH binary accretes at the same rate as a single massive BH, implying that binaries remain bright even near merger, with periodic modulations of their light-curves~\cite{Farris:2014zjo,Tang:2018rfm}. Typical massive BH binaries can be localized in the sky about a day (or 10-20 orbits) before merger, allowing a targeted search for periodic counterparts. Additionally, post-merger afterglows may be detectable~\cite{Schnittman:2008ez,Lippai:2008fx,Rossi:2009nk,Anderson:2009fa}. With the availability of 3G ground based GW detectors, EM counterparts from binary neutron star and neutron star-BH binaries may be observable up to $z=1$~\cite{Chen:2020zoq}, potentially enabling a sub-percent measurement of the Hubble constant. Multimessenger observations at such large redshifts require telescopes that can cover the full sky with large cadence and with the ability to detect faint sources. With the advent of large all-sky time-domain surveys, such as the Legacy Survey of Space and Time (LSST) by the Vera Rubin Observatory~\cite{LSSTScience:2009jmu}, a large fraction of the sky will have archival data on faint sources at high cadence (at least weekly), allowing archival searches for EM counterparts up to high redshift.

\noindent{\bf \em Dark standard sirens:} Dark standard sirens are GW sources such as stellar origin BH binaries, or binary neutron star and neutron star-BH systems without detectable EM counterparts. The redshift of these sources cannot be identified from the host galaxy (see, however, Ref.~\cite{Borhanian:2020vyr} for how exceptional BH mergers could localize a single host galaxy within its localization patch) but it may be identified using either statistical host identification techniques~\cite{Schutz:1986gp,DelPozzo:2012zz,Chen:2017rfc,LIGOScientific:2018gmd,Gray:2019ksv} or by exploiting the spatial clustering between GW sources and galaxies to infer the clustering redshift of the GW sources through cross-correlation techniques~\cite{Oguri:2016dgk,Mukherjee:2018ebj,Mukherjee:2019wfw,Mukherjee:2020hyn,Bera:2020jhx,Mukherjee:2022afz}. By cross-correlating with spectroscopic or photometric galaxy surveys~\cite{Diaz:2021pem}, dark standard sirens can be used to measure the value of the Hubble constant up to redshift $z\sim 1$ with a few percent precision, even if the host galaxy of the GW source is not present in the galaxy catalog.  Upcoming galaxy surveys such as DESI~\cite{Aghamousa:2016zmz}, SPHEREx~\cite{Dore:2014cca,Dore:2018kgp}, Euclid~\cite{Refregier:2010ss}, Vera Rubin Observatory~\cite{LSSTScience:2009jmu} and the Roman Space Telescope~\cite{Dore:2018smn} will enable a nearly full sky measurement of the galaxy distribution with photometric/spectroscopic redshift measurements~\cite{Diaz:2021pem}.  The cross-correlation technique and the statistical host identification will also be useful for the dark standard sirens detectable from LISA in combination with the Vera Rubin Observatory, which can observe EM counterparts at high redshift.  In the future, cross-correlation of 3G detector measurements with the CMB, 21 cm and other line intensity signals will be useful to map the expansion history up to high redshift~\cite{Mukherjee:2019wfw,Scelfo:2020jyw,Scelfo:2021fqe}.

\noindent{\bf \em Spectral standard sirens:} An alternative to bright or dark sirens is to use information about the intrinsic spectrum of GW source properties to provide a scale, and thereby calibrate the signal and allow for the inference of redshift. The most straightforward approach to this would be to use the mass spectrum of GW events to measure redshift~\cite{Chernoff:1993th,Farr:2019twy,Ezquiaga:2020tns,Mastrogiovanni:2021wsd,LIGOScientific:2021aug,Ezquiaga:2022zkx}. This method has been implemented on the O3 LIGO/Virgo/KAGRA data, providing $\sim15\%$ constraints on $H_0$~\cite{LIGOScientific:2021aug}, and is expected to provide percent-level constraints at $z\gtrsim1$ with current and future GW detector networks~\cite{Taylor:2011fs,Farr:2019twy,Ezquiaga:2020tns,You:2020wju,Mastrogiovanni:2021wsd,Ezquiaga:2022zkx} after mitigating the effects of formation scenarios and dependence on stellar metallicity~\cite{Mukherjee:2021rtw}. An alternative approach is to use the EoS of neutron stars to provide a unique signature in the mass distribution, and thereby infer redshift and constrain cosmology using binary neutron stars as standard sirens in the absence of EM counterparts~\cite{Messenger:2011gi,Chatterjee:2021xrm}.

\subsection{Dark energy and cosmic expansion rate at higher redshift}

High-redshift ($z\gtrsim 0.1$) GW events offer additional opportunities to test the presence of dark energy. To measure the dark energy equation of state $w(z)$ we need simultaneous measurements of the luminosity distance and counterpart redshift of GW sources, since the relationship between these two numbers depends on $w(z)$. This may be possible by cross-correlating GW sources with the galaxy distribution identified from photometric or spectroscopic redshift surveys~\cite{Mukherjee:2020hyn}, or by identifying host galaxies from concomitant EM signals~\cite{Cutler:2009qv,Tamanini:2016zlh,Oguri:2016dgk,Mukherjee:2020hyn,Diaz:2021pem}.

The measurements of GW sources up to high redshifts allowed by LISA ($z\lesssim 10$), Einstein Telescope and Cosmic Explorer ($z\lesssim 1$ for binary neutron stars and $z\lesssim 80$ for binary BHs) will map the cosmic expansion rate well beyond redshift $z=1$.  Therefore GW sources observed by next-generation detectors will probe the properties of dark energy in uncharted territory. GW sources will also be able to reconstruct the expansion rate of the Universe up to high redshift in a model-independent way, independently validating the different components that contribute to the energy budget of the Universe.

\clearpage
\section{Cosmological gravitational waves}
\label{sec:cosmo}

\begin{figure}[t]
\centering
\includegraphics[width=0.6\textwidth]{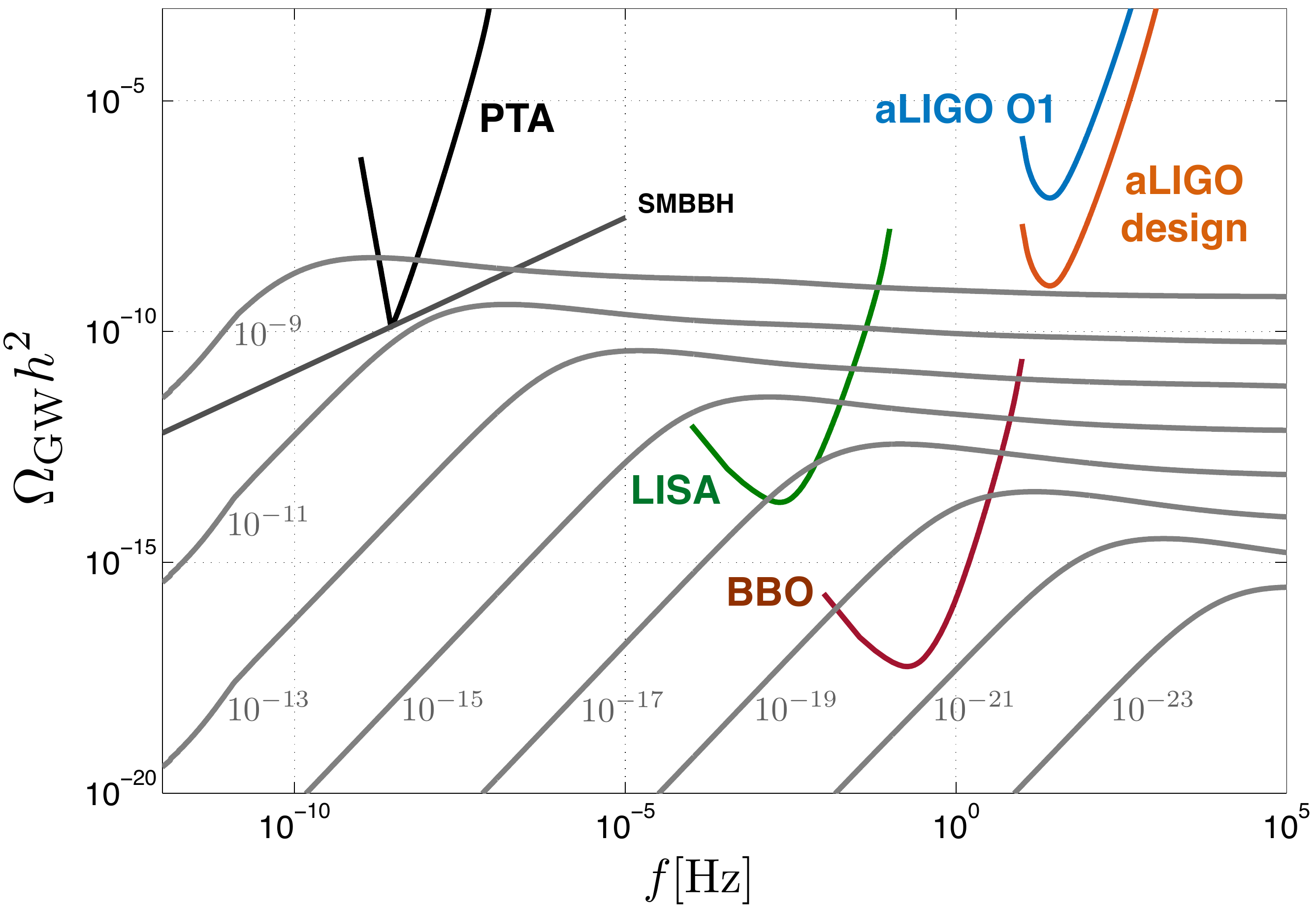}
\caption{Cosmic (super)string GW spectra for values of the dimensionless string tension  $G\mu/c^2$ in the range of $10^{-23}$-$10^{-9}$, as well as the spectrum produced by supermassive BH binaries, along with current and future experimental constraints. PTA sensitivity will not be superseded until the LISA mission. The Big Bang Observer (BBO) is a future planned space-based GW detector. Figure from Ref.~\citep{Blanco-Pillado:2017rnf}.}
\label{fig:omega}
\end{figure}

\subsection{Cosmic (super)strings}

Cosmic strings are topological defects that can form during phase transitions in the early Universe~\citep{Kibble:1976sj,Vilenkin:2000jqa}, and cosmic superstrings are the fundamental strings of string theory stretched to cosmological scales due to the expansion of the Universe~\citep{Jones:2002cv,Sarangi:2002yt,Dvali:2003zj,Jones:2003da,Copeland:2003bj,Jackson:2004zg}. In a cosmological setting, and for the simplest superstring models, cosmic string and superstring networks evolve in the same way. For a detailed review of cosmic (super)string network evolution and observational signatures, see e.g.~\citep{Copeland:2009ga}. Cosmic (super)strings can exchange partners when they meet and produce loops when they self-intersect. These loops then oscillate and lose energy to GWs generating bursts and a stochastic background~\citep{Berezinsky:2001cp,Damour:2000wa,Damour:2001bk,Damour:2004kw,Siemens:2006vk,Siemens:2006yp}: these are the signals we wish to detect~\citep{Pastorello:2019akb,Blanco-Pillado:2017rnf}. Strings are characterized by their mass per unit length $\mu$, which is normally given in terms of the dimensionless parameter $G\mu/c^2$, the ratio of the string energy scale to the Planck scale squared. The detection of a stochastic background from cosmic (super)strings, or GWs from individual cosmic (super)string loops, would be transformative for fundamental physics.

\vspace{0.3cm}
\noindent
The cosmic string GW spectrum is broad-band, spanning many orders of magnitude in frequency, and it is therefore accessible to a number of GW experiments including LIGO, LISA, and PTAs. However, PTAs are currently the most sensitive experiment for the detection of cosmic (super)strings and will remain so for at least the next decade and a half. Correspondingly, pulsar-timing experiments are producing the most constraining bounds on the energy scale and other model parameters of cosmic strings and superstrings. The best limit on the string tension, $G\mu/c^2 < 5.3(2) \times 10^{-11}$, is several orders of magnitude better than constraints from cosmic microwave background (CMB) data, and comes from the NANOGrav Collaboration~\citep{NANOGRAV:2018hou}. Figure~\ref{fig:omega} shows the stochastic background spectrum produced by cosmic strings in terms of the dimensionless density parameter $\Omega$ versus frequency for dimensionless string tensions $G\mu/c^2$ in the range $10^{-23}$--$10^{-9}$. Overlaid are current and future experimental constraints from PTAs, ground-based GW detectors, and spaced-based detectors. As we mentioned, PTAs might have already seen the first hints of a stochastic background~\cite{NANOGrav:2020bcs}, and the implications of these hints for cosmic strings have been explored by many authors, see e.g.~\cite{Blanco-Pillado:2021ygr,Blasi:2020mfx,Afzal:2022vjx,Cyr:2022urs,Emond:2021vts, Buchmuller:2021mbb}. The PTA sensitivity to stochastic backgrounds of cosmic strings will not be superseded until the LISA mission, which is scheduled for launch in the mid 2030s.

\subsection{Primordial gravitational waves from inflation}

The evolution of the very early Universe is thought to include a period of exponential expansion called inflation that accounts for its observed homogeneity, isotropy, and flatness~\citep{Brout:1977ix,Starobinsky:1980te,Kazanas:1980tx,Sato:1980yn,Guth:1980zm,Linde:1981mu,Albrecht:1982wi}. Additionally, by expanding quantum fluctuations present in the pre-inflationary epoch, inflation seeds the density fluctuations that evolve into the large scale structures we see in the Universe today~\citep{Mukhanov:1981xt,Hawking:1982cz,Guth:1982ec,Starobinsky:1982ee,Bardeen:1983qw}, and produces a stochastic background of GWs~\citep{Starobinsky:1979ty,Rubakov:1982df,Abbott:1984fp}. This GW background is broad-band, like the one produced by cosmic strings, and potentially detectable by multiple experiments.

\vspace{0.3cm}
\noindent
Detecting primordial GWs from inflation has been a critical objective of CMB experiments for some time~\citep{Kamionkowski:2015yta}. The CMB is sensitive to the lowest frequency portion of the GW spectrum from inflation, and CMB data can be used to constrain the tensor-to-scalar ratio, which is the ratio of the size of GWs produced to that of scalar perturbations (which seed density fluctuations as described above).  For standard inflation models the GW background in the PTA band is likely to be fainter than that of supermassive BH binaries, though that depends in part on the character of the supermassive BH binary spectrum at the lowest frequencies, where environmental effects like accretion from a circumbinary disk or stellar scattering can reduce GW emission from supermassive BH binaries~\citep{Sampson:2015ada}.  In addition, some inflationary models have a spectrum that rises with frequency.  Thus, GW detectors operating at higher frequencies than CMB experiments, like PTAs and space- and ground-based interferometers, can be used to constrain the shape of the inflationary GW spectrum. Indeed, PTA, CMB, and GW interferometer data across 29 decades in frequency have already begun to place stringent limits on such models~\citep{Lasky:2015lej}.

\subsection{Gravitational waves from phase transitions}

The early Universe may have experienced multiple phase transitions as it expanded and cooled. Depending on the detailed physical processes that occur during a phase transition, GWs can be generated with wavelengths of order the Hubble length at the time of the phase transition. That length scale, suitably redshifted, translates into a GW frequency today. Thus, GW experiments at different frequencies today probe horizon-sized physical processes occurring at different times in the early Universe, with higher frequency experiments probing earlier and earlier times.

\begin{figure}[t]
\centering
\includegraphics[width=0.6\textwidth]{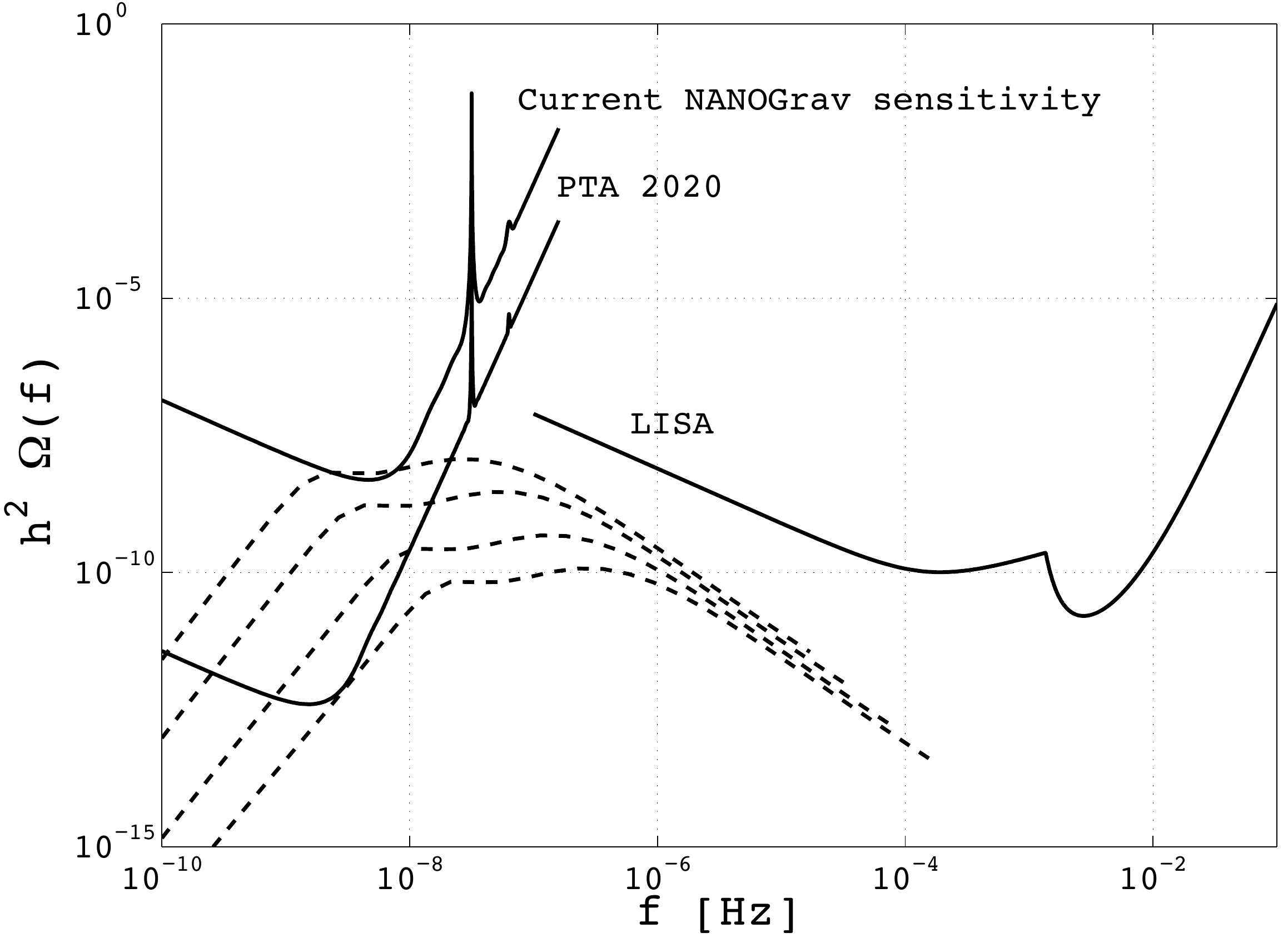}
\caption{GW spectrum of a first order QCD phase transition for various phase transition durations (dashed lines), along with the PTA and LISA sensitivities (solid lines). LISA will not be able to detect a signal from the QCD phase transition; the electroweak phase transition that occurs at higher temperatures, earlier times, and therefore maps onto higher frequencies, is a more promising source for LISA. Figure from Ref.~\citep{Caprini:2010xv}.}
\label{fig:qcd}
\end{figure}

\noindent
For example, the nanohertz frequency band accessible to PTAs maps onto the era in the early Universe when the quantum chromodynamics (QCD) phase transition took place, about $10^{-5}$~s after the Big Bang. The horizon at that time was on the order of 10~km, and any GWs generated at that length scale at that time would today be stretched to about 1~pc (or 3 light-years), which corresponds to GW frequencies of about 10~nHz, and lie within the PTA sensitivity band.  The possibility that interesting QCD physics can result in a GW signal detectable by PTAs was first pointed out by Witten in the 1980s~\citep{PhysRevD.30.272}.  More recently, Caprini et al.~\citep{Caprini:2010xv} considered the possibility of a first order phase transition at the QCD scale. In standard cosmology the QCD phase transition is only a cross-over, and we do not expect it to generate GWs. However, if the neutrino chemical potential is sufficiently large it can become first order (it is worth pointing out that if sterile neutrinos form the DM, we expect a large neutrino chemical potential).  There is also the possibility that the fluctuations of gluon fields could generate scalar GWs from the conformal anomaly in the quark-gluon plasma phase~\cite{Mottola:2016mpl}. Thus PTAs provide a window onto physical processes occurring in the Universe at the time of the QCD phase transition or before, and could detect GWs from a first order phase transition at that time (see Fig.~\ref{fig:qcd}). As can be seen in Fig.~\ref{fig:qcd}, a given PTA will become sensitive to lower frequencies as the baselines on their data sets increases through the next decade.

It should be noted that for PTAs all of these stochastic background signals (cosmic strings, inflation, and phase transitions) will have to contend with the background of supermassive BH binaries expected to be detected in the next few years by PTAs. However, long enough baselines and  sufficiently distinguishable spectral characteristics will make these signatures individually resolvable \citep{Parida:2015fma}. 

\section{Executive summary}

We conclude with a list of {\bf key opportunities} in the various sub-areas listed above:

\vspace{.5cm}

\noindent
{\bf Tests of strong-field gravity:}

\begin{itemize}
    \item Coordinate efforts in the development of the combined science case for future GW detectors, including 3G ground-based interferometers, space-based interferometers, atom interferometry and PTAs.
    \item Coordinate synergies between multimessenger and GW observations on the ground and in space.
    \item Explore connections between astrophysical tests of GW production and propagation and laboratory tests of gravity.
    \item Search for additional GW polarizations in binary mergers and in the cosmological stochastic GW background using PTAs and the expanding network of GW interferometers, which could hint to modified gravity.
\end{itemize}

\noindent
{\bf Black hole horizons, quantum gravity and the information paradox:}

\begin{itemize}
    \item Explore connections between EM tests of the Kerr metric (based e.g. on EHT and GRAVITY observations) and GW tests of BH dynamics.
    \item Construct ``ab initio'' models of nonsingular, horizonless alternatives to BHs, and self-consistent predictions of the ringdown spectra and echo signal they would produce.
    \item Improve theory and data analysis methods for BH spectroscopy, as well as searches for horizonless compact objects (using e.g. echoes or tidal deformability measurements) in the large sample of large-SNR sources observable by next-generation GW detectors.
\end{itemize}

\noindent
{\bf Gravitational signatures of dark matter:}

\begin{itemize}
    \item Look for smoking-gun signatures of dark compact objects, including e.g. accretion-induced collapse and their effects in gravitational waveforms.
    \item Investigate gravitational signatures of dilute dark matter distributions on binary dynamics and/or PTA signals.
    \item Search for nonperturbative signatures of ultralight bosonic fields, such as gaps in their astrophysical mass/spin distribution, stochastic or resolvable signals produced by BH/boson cloud systems, bosenovas, EM signals due to axion/photon couplings, and the effect of boson clouds on stellar distributions around massive BHs.
\end{itemize}

\noindent
{\bf Expansion history of the Universe using multi-messenger observations:}

\begin{itemize}
    \item Exploit the wealth of information coming from bright and dark GW standard sirens detectable by ground- and space-based detectors to map the cosmic expansion history up to high redshift.
    \item Coordinate GW detectors and EM observations with large sky coverage and cadence over multiple frequency bands to open up new avenues in multimessenger astronomy, addressing several open problems in astrophysics and fundamental physics.
\end{itemize}

\noindent
{\bf Cosmological gravitational waves:}

\begin{itemize}
    \item Use the network of existing and future GW detectors (ground-based interferometers, space-based interferometers, atom interferometers, and PTAs) to search for GW signals from cosmological sources including cosmic strings and superstrings, phase transitions in the early Universe, and inflation.
\end{itemize}

\clearpage

\section*{List of endorsers}

Miguel \'A. Aloy (University of Valencia)\\
Lorenzo Annulli (University of Aveiro)\\
Marie-Christine Angonin (SYRTE, Observatoire de Paris, Université PSL, CNRS, Sorbonne Université, LNE)\\
Andrea Antonelli (Johns Hopkins University)\\
K. G. Arun (Chennai Mathematical Institute, India)\\
Vishal Baibhav (Northwestern University)\\
Sambaran Banerjee (University of Bonn)\\
Elvis Barakovic (University of Tuzla)\\
Enrico Barausse (SISSA)\\
Barry C. Barish (Caltech \& UC Riverside) \\
Daniel Baumann (Amsterdam \& National Taiwan University)\\
Marc Besancon (CEA Paris-Saclay University, Irfu, DPhP)\\
Jiri Bicak (Charles University Prague)\\
Ofek Birnholtz (Bar-Ilan University)\\
Marie Anne Bizouard (Artemis, Observatoire de la Cote d'Azur)\\
Diego Blas (Universitat Aut\`onoma de Barcelona \& IFAE)\\
Jose Luis Blázquez-Salcedo (Complutense University of Madrid)\\
Alexey Bobrick (Technion, Israel Institute of Technology)\\
Tamara Bogdanovi\'c (Georgia Institute of Technology)\\
B\'eatrice Bonga (Radboud University, Nijmegen)\\
Christian G. B\"ohmer (University College London)\\
Alexander Bonilla (Universidade Federal de Juiz de Fora)\\
Elisa Bortolas (Universit\`a degli Studi di Milano-Bicocca)\\
Pasquale Bosso (University of Lethbridge)\\
Philippe Brax (IPhT Paris-Saclay University)\\
Richard Brito (CENTRA, Instituto Superior T\'{e}cnico, Portugal)\\
Marco Bruni (University of Portsmouth)\\
Duncan A.~Brown (Syracuse University)\\
Tomasz Bulik (University of Warsaw)\\
Laura Cadonati (Georgia Tech)\\
Robert Caldwell (Dartmouth College)\\
Enrico Calloni (University Federico II,  Naples, Italy)\\
Pedro R. Capelo (University of Zurich)\\
Marco Cavagli\`a (Missouri University of Science and Technology)\\
Srija Chakraborty(Scuola Normale SUperiore, Pisa, Italy)\\
Naz{\i}m \c{C}abuk (Ankara University)\\
Senem \c{C}abuk (Ankara University)\\
Mesut \c{C}al{\i}\c{s}kan (Johns Hopkins University)\\
Alejandro C\'ardenas-Avenda\~no (Princeton University \& Konrad Lorenz University)\\
Cecilia Chirenti (University of Maryland)\\
Giacomo Ciani (University of Padova, Padua, Italy)\\
Katy Clough (Queen Mary, University of London)\\
Lucas G. Collodel (University of T\"ubingen)\\
Alessandra Corsi (Texas tech University)\\
Djuna Croon (IPPP Durham)\\
Yanou Cui (University of California-Riverside)\\
Giulia Cusin (University of Geneva \& Institut d'Astrophysique de Paris)\\
Saurya Das (University of Lethbridge)\\
Jordy Davelaar (Columbia University \& Flatiron Institute)\\
Benjamin L. Davis (New York University Abu Dhabi)\\
Mariafelicia De Laurentis (University of Naples "Federico II", Naples, Italy)\\
Valerio De Luca (University of Geneva) \\
Andrea Derdzinski (University of Zurich)\\
Kyriakos Destounis (University of T\"ubingen)\\
Valerie Domcke (CERN/EPFL)\\
Daniela Doneva (University of T\"ubingen)\\
Amelia Drew (University of Cambridge)\\
Francisco Duque (CENTRA, IST, Universidade de Lisboa)\\
Ruth Durrer (Geneva University)\\
Stephanie Escoffier (Aix Marseille Univ, CNRS/IN2P3, CPPM)\\
Jose Maria Ezquiaga (University of Chicago)\\
Hontas F. Farmer (Elmhurst University \& College of DuPage)\\
Will M. Farr (Stony Brook University \& CCA)\\
Pedro G. Ferreira (University of Oxford)\\
Scott Field (University of Massachusetts Dartmouth)\\
Daniel G.~Figueroa (IFIC, Valencia, Spain)\\
Stanislav Fisenko (Moscow State Linguistic University)\\
Giacomo Fragione (Northwestern University)\\
Gabriele Franciolini (Sapienza University of Rome)\\
Jacopo Fumagalli (IFT UAM-CSIC, Madrid)\\
Mudit Garg (University of Zurich)\\
Edgar Gasper\'in (CENTRA, Instituto Superior T\'ecnico)\\
Suvi Gezari (Space Telescope Science Institute)\\
Mandeep S. S. Gill (Stanford University)\\
Sarah E. Gossan (Canadian Institute for Theoretical Astrophysics)\\
Luca Graziani (Sapienza, University of Rome)\\
Leonardo Gualtieri (University of Rome ``La Sapienza'')\\
Anuradha Gupta (The University of Mississippi)\\
Daryl Haggard (McGill University)\\
Wen-Biao Han (Shanghai Astronomical Observatory, CAS)\\
Troels Harmark (Niels Bohr Institute, University of Copenhagen)\\
Gregory Harry (American University, Washington DC)\\
Aurelien Hees (SYRTE, Observatoire de Paris, Universit\'e PSL, CNRS, Sorbonne Universit\'e, LNE)\\
Xunyang Hong (ETH Zürich)\\
Bala Iyer (ICTS-TIFR, Bangalore, India)\\
Rajeev Kumar Jain (Indian Institute of Science, Bangalore, India)\\
Philippe Jetzer (University of Z\"urich)\\
Shang-Jie Jin (Northeastern University, China)\\
Cristian Joana (University of Louvain)\\
Panagiota Kanti (University of Ioannina)\\
Bradley J.~Kavanagh (IFCA, UC-CSIC, Spain)\\
Fech Scen Khoo (University of Oldenburg)\\
Claus Kiefer (University of Cologne)\\
Masashi Kimura (Rikkyo University)\\
Antoine Klein (University of Birmingham)\\
Sergei A. Klioner (Technische Universit\"at Dresden)\\
Kostas Kokkotas (University of T\"ubingen)\\
Shimon Kolkowitz (University of Wisconsin - Madison)\\
Joachim Kopp (CERN \& JGU Mainz)\\
Ely D.~Kovetz (Ben Gurion University of the Negev)\\
Kevin Kuns (MIT)\\
Macarena Lagos (Columbia University)\\
Astrid Lamberts (Observatoire de la Côte d'Azur)\\
Philippe Landry (Canadian Institute for Theoretical Astrophysics)\\
Michele Lenzi (Institute of Space Sciences, ICE-CSIC and IEEC)\\
Christophe Le Poncin-Lafitte (SYRTE, Observatoire de Paris, Université PSL, CNRS, Sorbonne Université, LNE)\\
Dicong Liang (Peking University)\\
Chang Liu (Peking University)\\
Nicole Lloyd-Ronning (Los Alamos National Lab)\\
Giuseppe Lodato (Universit\`a degli Studi di Milano)\\
Francisco S. N. Lobo (University of Lisbon)\\
Thomas J. Maccarone (Texas Tech University)\\
Elisa Maggio (Max Planck Institute for Gravitational Physics, Albert Einstein Institute, Potsdam)\\
Maciej Maliborski (University of Vienna)\\
Szabolcs Marka (Columbia University in the City of New York)\\
Andrea Maselli (Gran Sasso Science Institute)\\
Lucio Mayer (University of Zurich)\\
Jurgen Mifsud (University of Malta)\\
Cole Miller (University of Maryland)\\
Chiara M. F. Mingarelli (University of Connecticut \& Flatiron Institute)\\
Hiroyuki Nakano (Ryukoku University)\\
Petya Nedkova (Sofia University)\\
David Neilsen (Brigham Young University)\\
David A.\ Nichols (University of Virginia)\\
Emil Nissimov (Institute for Nuclear Research and Nuclear Energy, Sofia)\\
Alexander H. Nitz (Max Planck Institute for Gravitational Physics, Albert Einstein Institute, Hannover)\\
Johannes Noller (University of Portsmouth \& University of Cambridge)\\
Rafael C. Nunes (Instituto Nacional de Pesquisas Espaciais, Brazil)\\
Giorgio Orlando (University of Groningen)\\
Richard O'Shaughnessy (Rochester Institute of Technology)\\
Fabio Pacucci (Center for Astrophysics $\vert$ Harvard \& Smithsonian)\\
Antonella Palmese (University of California Berkeley)\\
Rodrigo Panosso Macedo (University of Southampton)\\
Sohyun Park (CERN)\\
Vedad Pasic (University of Tuzla)\\
Hiranya V. Peiris (University College London \& Stockholm University)\\
Wlodzimierz Piechocki (National Centre for Nuclear Research)\\
Igor Pikovski (Stockholm University \& Stevens Institute of Technology)\\
Geraint Pratten (University of Birmingham)\\
David Radice (Penn State)\\
Guilherme Raposo (CENTRA, Instituto Superior Técnico \& CIDMA, University of Aveiro)\\
S\'ebastien Renaux-Petel (Institut d'Astrophysique de Paris, CNRS)\\
Jonathan W. Richardson (University of California, Riverside)\\
Keith Riles (University of Michigan)\\
Justin L. Ripley (University of Cambridge)\\
Liliana Rivera Sandoval (University of Texas Rio Grande Valley)\\
Marek Rogatko (Maria Curie Sklodowska University)\\
Dorota Rosinska (University of Warsaw)\\
Elena Maria Rossi (Leiden University)\\
Stephan Rosswog (Stockholm University)\\
Friedrich R{\"o}pke (Heidelberg University \& Heidelberg Institute for Theoretical Studies)\\
Diego Rubiera-Garcia (Complutense University of Madrid)\\
Javier Rubio (CENTRA, Instituto Superior T\'ecnico)\\
Milton Ruiz (University of Illinois at Urbana-Champaign)\\
Mairi Sakellariadou (King's College London)\\
Zeyd Sam (University of Potsdam)\\
Nicolas Sanchis-Gual (University of Valencia)\\
Misao Sasaki (University of Tokyo)\\
Lijing Shao (Peking University)\\
David Shoemaker (MIT)\\
Deirdre Shoemaker (University of Texas at Austin)\\
Jing Shu (CAS Key Laboratory of Theoretical Physics, Institute of Theoretical Physics, Chinese Academy of Sciences, Beijing 100190, P.R.China)\\
Alicia M. Sintes (Universitat de les Illes Balears\\
Joshua Smith (California State University, Fullerton)\\
Carlos F. Sopuerta (Institute of Space Sciences, ICE-CSIC and IEEC)\\
Lara Sousa (Instituto de Astrofísica e Ciências do Espaço, Universidade do Porto)\\
Ulrich Sperhake (University of Cambridge)\\
Ciprian Sporea (West University of Timisoara)\\
Oleksandr Stashko (Taras Shevchenko National University of Kyiv)\\
Leo C. Stein (University of Mississippi)\\
Chris Stevens (University of Canterbury)\\
Timothy J. Sumner (Imperial College London)\\
Ling Sun (The Australian National University)\\
Arthur G. Suvorov (Manly Astrophysics)\\
Tayebeh Tahamtan (Charles University)\\
Tomas Tamfal (University of Zurich)\\
David B. Tanner (University of Florida)\\
Thomas M. Tauris (Aalborg University)\\
Marika Taylor (University of Southampton)\\
Stephen R.~Taylor (Vanderbilt University)\\
Alejandro Torres-Orjuela (Sun Yat-Sen University)\\
Elias C. Vagenas (Kuwait University)\\
Alex Vano-Vinuales (CENTRA, IST, University of Lisbon)\\
John Veitch (University of Glasgow)\\
Jos\'e Velhinho (Beira Interior University, Portugal)\\
Vincent Vennin (APC, Paris University)\\
Daniele Vernieri (University of Naples and INFN Sezione di Napoli, Italy)\\
Flavio Vetrano (Urbino University, Italy)\\
Rodrigo Vicente (IFAE, Spain)\\
Matteo Viel (SISSA, Scuola Internazionale Studi Superiori Avanzati, Trieste , Italy)\\
Vincenzo Vitagliano (University of Genova)\\
Jeremy M.~Wachter (Skidmore College)\\
Barry Wardell (University College Dublin)\\
Alan J.~Weinstein (Caltech)\\
Toby Wiseman (Imperial College London)\\
Helvi Witek (University of Illinois at Urbana-Champaign)\\
Vojt{\v e}ch Witzany (University College Dublin)\\
Kinwah Wu (MSSL, University College London)\\
Fei Xu (University of Chicago)\\
Kent Yagi (University of Virginia)\\
Kadri Yakut (University of Ege)\\
Stoytcho Yazadjiev (University of Sofia)\\
Jaiyul Yoo (University of Z\"urich)\\
Silvia Zane (MSSL, University College London)\\
Junjie Zhao (Beijing Normal University)\\
Valery Zhdanov (Taras Shevchenko National University of Kyiv)\\
Bei Zhou (Johns Hopkins University)\\
Miguel~Zilhão (University of Aveiro)\\

\clearpage
\bibliographystyle{JHEP.bst}
\bibliography{main}

\end{document}